\documentclass[twocolumn,english,pra,longbibliography]{revtex4-2}
\usepackage[T1]{fontenc}
\usepackage[utf8]{inputenc}
\usepackage{color,babel,amsmath,amssymb,esint,hyperref,soul,graphicx}

\makeatletter

\usepackage{braket}
\usepackage{booktabs,microtype}

\makeatother

\begin{document}
\title{Bogoliubov-picture thermofield dynamics in anharmonic environments}%: finite-temperature quantum dynamics in a Morse oscillator bath}
\author{Rocco Martinazzo}
\email{rocco.martinazzo@unimi.it}
\affiliation{Department of Chemistry, Universit\`a degli Studi di Milano, Via Golgi 19, 20133 Milano}
\author{Bartosz B{\l}asiak}
\affiliation{Institute of Physical and Theoretical Chemistry, Goethe University Frankfurt, Max-von-Laue-Str. 7, D-60438 Frankfurt/Main, Germany}
\author{Irene Burghardt}
\affiliation{Institute of Physical and Theoretical Chemistry, Goethe University Frankfurt, Max-von-Laue-Str. 7, D-60438 Frankfurt/Main, Germany}

\begin{abstract}
Thermofield dynamics (TFD) describes finite-temperature quantum relaxation through a purified wavefunction, without sampling thermally populated initial states. For harmonic environments, TFD is typically combined with a thermal Bogoliubov transformation to obtain a compact wavefunction representation. This transformation can also be useful for anharmonic baths, but preparing the corresponding thermal state in the transformed representation poses additional numerical challenges. Here, we explore TFD approaches tailored to anharmonic potentials and compatible with efficient tensor network methods for quantum propagation. We show that the conceptually appealing route of converting a harmonic-reference thermal state into the target anharmonic state by imaginary-time propagation can be numerically ill-conditioned. As a practical alternative, we formulate a multilayer multiconfiguration time-dependent Hartree (ML-MCTDH) protocol that separates thermal-state preparation from the harmonic-reference Bogoliubov transformation. We prepare the thermal purification in the bare representation and then apply finite-basis Bogoliubov unitaries to the physical--auxiliary mode pairs. For independent bath modes, direct construction of the thermal amplitude provides a complementary initialization procedure. We illustrate this protocol through vibrational relaxation and spatial-superposition decoherence of a harmonic subsystem coupled to a Morse-oscillator bath. We find non-monotonic energy relaxation, shifts in the dominant resonant response, and oscillatory energy exchange at zero and finite temperature. For decoherence, we recover the expected harmonic-bath scaling and find a pronounced non-monotonic dependence of the decay rate on anharmonicity at finite temperature in a low-frequency model, with a weaker dependence in a high-frequency model.

\end{abstract}
\maketitle

\section{Introduction}

Temperature influences molecular quantum dynamics in condensed phases through the populations and fluctuations of the environment with which a molecular subsystem exchanges energy \cite{Nitzan2006,May2011,Breuer2007,Weiss2008}. Molecular excitations can relax into an already thermally populated environment, while thermal fluctuations can promote excitations and activate processes that are otherwise suppressed \cite{Nitzan2006,May2011}. Superpositions of subsystem states can lose coherence even when energy transfer is small, and their coherence lifetimes can depend strongly on temperature \cite{Schlosshauer2007}. Describing these processes on the same footing requires both a well-defined thermal initial state and a dynamical treatment capable of retaining system--environment correlations.

The open-system problem can be approached through a closed-system description containing an explicit, sufficiently large model of the environment \cite{Nitzan2006,May2011,Breuer2007,Weiss2008,Alonso2017,breuer2016,Xu2026}. The mixed-state character of the initial condition then poses an additional computational challenge. Wavepacket sampling represents thermal expectation values through an ensemble of pure-state propagations \cite{manthe2001,Gelman2003,WangSkinnerThoss2006,WangThoss2006}. Random-phase thermal wavefunctions provide one such construction: imaginary-time filtering supplies Boltzmann weights, and averaging over random phases recovers the thermal trace \cite{Gelman2003,Nest2007,lorenza2014}. Individual trajectories can exploit efficient wavepacket algorithms and can be propagated independently, but each observable requires statistical convergence. Sampling uncertainty can be particularly inconvenient when resolving small differences between nearby parameter values or physical conditions. 

Alternatively, Liouville-space propagation evolves the density operator directly according to the Liouville--von Neumann equation. Vectorization casts this equation as a Schr\"odinger-like equation in a doubled space \cite{Suzuki1991,Zwanzig2001,Campaioli2024}. This avoids sampling, although the density operator and the action of its dynamical generator must still be represented efficiently.

Thermofield dynamics (TFD)\cite{Takahashi1975,umezawa_matsumoto_tachiki_1982,Arimitsu1985,Suzuki1985,Barnett1987,Suzuki1991,Takahashi1996,Borrelli2016,Borrelli2021} also introduces an enlarged Hilbert space, but represents a purification of the density operator or, equivalently, vectorizes the square root of the latter, $\sqrt{\rho}$. Auxiliary, or tilde \cite{Takahashi1975}, degrees of freedom are introduced so that tracing them out recovers the physical density operator. Thermal averages then become ordinary expectation values in the purified state. 
The purified state evolves according to the physical Hamiltonian (up to a gauge choice) and can be propagated using wavefunction methods, including tensor network methods like matrix product states (MPS) \cite{White2004RealTimeDMRG,Verstraete2004MPSFiniteTemperature,LeDe2024MPSOpenSystems} as well as the multiconfiguration time-dependent Hartree (MCTDH) method \cite{mey90:73,man92:3199,bec00:1,mey09:book} and its multilayer extension (ML-MCTDH)\cite{Wang2003,ven11:044135}. 
The gauge freedom -- the possibility to rotate the auxiliary states -- is an added flexibility that can be exploited to enforce physical-auxiliary symmetries at the price of adding Hamiltonian terms for the auxiliary degrees of freedom. All this renders vectorization of the density matrix in Liouville space and its purification in TFD two fundamentally distinct constructions, even though both employ auxiliary states that are copies of physical states. 

For a harmonic environment, the thermal Bogoliubov transformation \cite{Barnett:85} is a unitary transformation that maps the thermal purification onto an uncorrelated oscillator vacuum, and that effectively introduces a change of picture of the dynamics \cite{Takahashi1996,Barnett:85,Gelin2017,Borrelli2021,blasiak2026}, referred to as "Bogoliubov picture" in the following. The transformed initial state is consequently particularly compact and extremely simple to prepare. Compactness generally translates to the dynamical state, though in a system-specific way, as it depends on how system--bath interactions generate excitations and correlations during propagation. This is a key advantage of the Bogoliubov transformation that effectively cools down the dynamics at $T=0$~K by introducing temperature-dependent interactions.

Anharmonic environments \cite{Wang2004AnharmonicBath,Makri1994AnharmonicBath,Makri2024AnharmonicBath,Rhee2025AnharmonicBath} may also benefit from the Bogoliubov transformation, which is a well-defined unitary transformation based on a harmonic reference. This can be beneficial when the transformed thermal state remains close to the reference vacuum, and the wavefunction compactness mentioned above can be transferred to the anharmonic environment. This possibility motivates preparing anharmonic thermal states directly in the Bogoliubov picture. 
Our approach differs from that of Gelin and Borrelli~\cite{Gelin2017}, who proposed generalized thermal transformations beyond harmonic oscillators, constructed in finite-level eigenstate representations. Here, we retain the harmonic-reference Bogoliubov transformation and prepare the anharmonic thermal amplitude independently, separating thermal-state preparation from the picture change.

A straightforward way to obtain the thermal purification of a bath Hamiltonian $H_B$ from that of a reference Hamiltonian $H_B^0$ is imaginary-time propagation with $K=H_B-\widetilde H_B^0$. We call this construction \emph{reference thermalization}, as opposed to the \emph{bare} thermalization \cite{Venrell2026} that starts from the infinite-temperature thermal state common to every Hamiltonian. The auxiliary factor $e^{+\beta\widetilde H_0/2}$ that appears in the former (see Section \ref{sec:imaginary-time preparation}) formally removes the reference thermal weights, while $e^{-\beta H_B/2}$ supplies those of the target Hamiltonian. This conversion does not require a Bogoliubov transformation, although it is conveniently performed in the Bogoliubov picture if the aim is to find the thermal vacuum of $H_B$ in this picture. However, it turns out that the inverse imaginary-time factor amplifies errors in high-energy auxiliary components, making the construction numerically ill-conditioned. Enlarging the underlying basis can therefore degrade, rather than improve, the prepared state. Dissociative potentials (like those analyzed in the following) pose a separate, additional problem: an isolated dissociative oscillator does not have a normalizable canonical state on an infinite coordinate domain, so the intended initial ensemble and its numerical preparation must be carefully specified.

The present work addresses these issues for the systematically tunable Morse-oscillator bath introduced in Ref.~\cite{MorseBath2011}. We describe two alternatives to reference thermalization: direct pair construction for independent bath oscillators (DC) and an apply-after-thermalizing procedure (AA), in which a Bogoliubov transformation follows bare thermalization. Both exploit the factorization of the harmonic-reference transformation into physical--auxiliary pairs. Each pair unitary is evaluated as a finite-dimensional matrix exponential and applied within a combined ML-MCTDH mode. Bare thermalization in AA can, in principle, also include interactions between bath modes before the local pair transformations are applied. We then compare vibrational relaxation from two subsystem preparations and examine decoherence of spatial superpositions, using in both cases a harmonic subsystem. The emphasis is on separating thermal-preparation errors from representation and propagation errors, and on identifying robust signatures of bath anharmonicity in the subsequent dynamics. 

All wavefunction propagations reported in this work were performed using the Heidelberg MCTDH package (version 8.6.10)~\cite{mctdh:MLpackage}, employing its multilayer implementation~\cite{ven11:044135}.

\section{Thermofield dynamics}
\subsection{Purification and auxiliary freedom}\label{Purification and TFD}

In TFD, the purified state is defined in the doubled Hilbert space $\mathcal{H}_D=\mathcal{H}\otimes\widetilde{\mathcal{H}}$,
where $\mathcal{H}$ is the Hilbert space of the quantum mechanical system of interest and $\widetilde{\mathcal{H}}$, the auxiliary or tilde space, is an anti-linear copy of it. 
This space $\mathcal{H}_D$ is essentially the same Hilbert space used to represent (Hilbert-Schmidt) operators in Liouville dynamics as vectors. Vectorization is formally simplified by the introduction of the (improper) unit vector 
\begin{equation}
 \ket{I}=\sum_n\ket{n}\ket{\widetilde n}
\end{equation}
that represents the identity operator in an arbitrary basis $\{\ket{n}\}$ and its tilde image $\{\ket{\tilde{n}}\}$, and that encodes the physical-tilde mapping,
\begin{equation}
T:\mathcal{H}\rightarrow \widetilde{\mathcal{H}},\,\,\, \ket{\tilde{n}}=T\ket{n}:=\braket{n|I}
\end{equation}
(and similarly, $\braket{\tilde{n}|I}=\ket{n}$). Every system operator $A$ can then be vectorized according to $\ket{A}:=(A\otimes 1)\ket{I}$. In Liouville dynamics density operators $\rho$ are represented as $\ket{\rho}$ and ensemble averages take the form of scalar products in the enlarged space, $\braket{A}=\braket{\rho|A}$.  
Alternatively, TFD defines \emph{purified} density operators according to 
%Alternatively, and this is TFD, density operators are \emph{purified} according to 
$\ket{\Psi_\rho}=(\Gamma_\rho\otimes1)\ket{I}$, where $\Gamma_\rho$ is a (generalized) square root of $\rho$, i.e., it is such that $\Gamma_\rho\Gamma_\rho^\dagger=\rho$. In this way, ensemble averages become ordinary expectation values in the doubled space, 
\[\braket{A}=\braket{\Psi_\rho |A|\Psi_\rho},\] 
since $\rho\equiv \widetilde{\textrm{Tr}}\left(\ket{\Psi_\rho}\bra{\Psi_\rho}\right)$, where the trace is over the auxiliary space. For instance, 
\begin{equation}
 \ket{0(\beta)}=\frac{1}{\sqrt{Z}}(e^{-\beta H/2}\otimes 1)\ket{I}\,\,\textrm{with}\,\,\,Z=\textrm{Tr} e^{-\beta H}
 \label{eq:purification}
\end{equation}
represents the purified thermal equilibrium state at inverse temperature $\beta$ for a system described by the Hamiltonian $H$. For such an equilibrium state physical observables and their tilde copies have identical expectation values
\begin{equation}
\begin{aligned}
\braket{0(\beta)|A|0(\beta)}
&= \frac{1}{Z_\beta}
   \braket{I|e^{-\beta H/2}e^{-\beta \tilde{H}/2}A|I}\\
&= \frac{1}{Z_\beta}
   \braket{I|e^{-\beta H/2}e^{-\beta \tilde{H}/2}
   \tilde{A}^{\dagger}|I}\\
&= \frac{1}{Z_\beta}
   \braket{I|e^{-\beta \tilde{H}/2}
   \tilde{A}^{\dagger}e^{-\beta H/2}|I}\\
&= \braket{0(\beta)|\tilde{A}^{\dagger}|0(\beta)}\label{eq:thermal condition}
\end{aligned}
\end{equation}
where $\tilde{A}=T A T^{-1}$, and where we have repeatedly used the identity $O\ket{I}=\tilde{O}^\dagger\ket{I}$ for $O=A,H$ and simplified the notation according to $O\equiv O\otimes 1$ and $\tilde{O}\equiv 1\otimes \tilde{O}$. The identity $O\ket{I}=\tilde{O}^\dagger\ket{I}$ is very useful in practice and introduces the so-called weak equivalence \cite{Suzuki1991} $O\sim\tilde{O}^\dagger$. Of course, $O\sim \tilde{O}$ when $O=O^\dagger$, and in Eq.~\eqref{eq:thermal condition} we can replace $\tilde{A}^\dagger$ with $\tilde{A}$ when $A$ is an observable. As we shall see later, Eq.~\eqref{eq:thermal condition} represents a useful diagnostics for the quality of a numerically computed equilibrium state.

The purified state evolves in time according to the same time-dependent  Schr\"odinger equation governing the dynamics of a pure system state. Indeed, we can set
\begin{equation}
    \rho(t)=\Gamma_\rho(t)\Gamma_\rho(t)^\dagger\,\,\textrm{with}\,\,\,\Gamma_\rho(t)=U_t\Gamma_\rho(0),
\end{equation}
where $\Gamma_\rho(0)\Gamma_\rho(0)^\dagger=\rho(0)$ is the initial density operator and $U_t$ is the time-evolution operator generated by the Hamiltonian $H$. Therefore,
\begin{equation}
    \ket{\Psi_\rho(t)}=\Gamma_\rho(t)\ket{I}= U_t\Gamma_\rho(0)\ket{I}=U_t\ket{\Psi_\rho(0)}
\end{equation}
However, applying an auxiliary unitary leaves the physical density operator unchanged.
Therefore, an arbitrary tilde Hamiltonian contribution $\widetilde G$ can be added to $H$ without altering the dynamics of the physical subsystem. 
The common choice $\widetilde G=-\widetilde H$ (making the purified thermal state stationary) is convenient in connection to the Bogoliubov transformation, but other choices are possible. Since $H_\textrm{TFD}=H-\tilde{H}$ is also the double-space representation of the Liouville superoperator in Liouville dynamics, we call this choice the Liouville gauge. The bare gauge corresponds to $\widetilde G=0$, while the choice $\widetilde G=+\widetilde H$ can be called symmetric gauge.

Irrespective of the gauge choice, the tilde states merely play the role of `ancilla' states, dynamically uncoupled from the physical ones. 
Therefore, full doubling of the space may be unnecessary: 
for a factorized initial state with a pure subsystem, $\rho(0)=\ket{\psi_S}\bra{\psi_S}\otimes\rho_B$, only the bath requires purification. 
The initial thermofield state can be written as $\ket{\psi_S}\ket{0(\beta)}$, without introducing an auxiliary subsystem state $\ket{\tilde{\psi}_S}$ that would remain at any time a separated factor in the full wavefunction. Similar considerations apply to high-frequency modes ($\hbar \omega_k\gg k_B T$) that hardly occupy their first excited state and are thus well described by a pure state. 

The Bogoliubov transformation introduced in the following section, Sec.~\ref{sec:thermal BT}, mixes physical and auxiliary variables and generally introduces explicit couplings
between them. In the corresponding picture, both physical observables and auxiliary gauge generators must be
transformed consistently. The gauge freedom is preserved, but is generally no longer represented by rotations acting
on the auxiliary coordinates alone. For the mode-wise construction used here, any selection of modes to be
doubled is therefore made in the bare representation, before applying the transformation to the retained physical--auxiliary pairs.

%Notice, however, that when performing the Bogoliubov transformation discussed in next Section, in Sec. \ref{sec:thermal BT}), and changing to the corresponding picture, an explicit coupling results between physical and tilde variables from the change of representation. Therefore, the gauge freedom above involves transformed observables and is no longer of immediate use. The bottom line is that in this picture, any mode selection for doubling need to be performed \emph{before} applying such transformation.

\subsection{Thermal Bogoliubov transformation}\label{sec:thermal BT}
For independent oscillators with the reference Hamiltonian 
\begin{equation}
    H_B^0 = \sum_{k=1}^{N_B} \hbar\omega_k\left(a_k^\dagger a_k+\frac{1}{2}\right)
\end{equation}
the mode-wise Bogoliubov transformations
\begin{equation}
    \begin{aligned}
    b_k = \cosh(\theta_k) a_k -\sinh(\theta_k) \tilde{a}^\dagger_k\\
    \tilde{b}_k =-\sinh(\theta_k) a^\dagger_k + \cosh(\theta_k) \tilde{a}_k 
    \end{aligned}
\end{equation}
preserve the bosonic commutation relations. With $\tanh\theta_k=e^{-\beta\hbar\omega_k/2}$, the transformed
operators annihilate the thermal equilibrium state $\ket{0_0(\beta)}$ 
\[
b_k\ket{0_0(\beta)}=\tilde{b}_k\ket{0_0(\beta)}=0
\]
The purified thermal state is therefore a vacuum for these operators, and for this reason is usually called \emph{thermal vacuum} \cite{Takahashi1996}.

The operator transformation is implemented in the extended Fock space by a unitary transformation
\begin{equation}
    b_k=U_\theta\, a_k\, U_\theta^\dagger\,\,\,\,\,\tilde{b}_k=U_\theta\, \tilde{a}_k\, U_\theta^\dagger \label{BT unitary}
\end{equation}
where $U_\theta$ takes the explicit form
\begin{equation}
 U_\theta=\prod_{k=1}^{N_B}\exp\!\left[\theta_k
 (a_k^\dagger\widetilde a_k^\dagger-a_k\widetilde a_k)\right].
 \label{eq:Utheta}
\end{equation}
This unitary can be conveniently used for changing the dynamical picture into one 
where the thermal state is represented by the \emph{bare} vacuum $\ket{\textbf{0}}$. 
With this convention $\ket{0_0(\beta)}=U_\theta\ket{\textbf{0}}$, and the transformed state and operators are, respectively,
\begin{equation}
    \ket{\Psi^\theta}=U_\theta^\dagger\ket{\Psi}\,\,\,A^\theta=U_\theta^\dagger A U_\theta\label{eq:iBT}
\end{equation}
We refer to $\ket{\Psi^\theta}$
and $A^\theta$ as states and operators in the Bogoliubov picture, noticing that $A^\theta$ 
is linked to $A$ via an \emph{inverse} Bogoliubov transformation ($i-$BT), 
see Eq. (\ref{BT unitary}) for comparison. For coordinates and momenta, for instance, the equations
\begin{align}
 q_k^\theta&=\cosh\theta_k\,q_k+\sinh\theta_k\,\widetilde q_k,\label{eq:BTq}\\
 p_k^\theta&=\cosh\theta_k\,p_k-\sinh\theta_k\,\widetilde p_k\label{eq: BTp}
\end{align}
allow one to represent the $i$-BT of the basic operators and of their functions in terms of ordinary, Schr\"odinger-picture ones. For a potential energy $V=V(q_1,q_2,..q_N)$, for example,
\begin{equation}
    V^\theta= V(..,q^\theta_k,..)=V(..,\cosh\theta_k\,q_k+\sinh\theta_k\, \tilde{q}_k,..)\label{eq:BTV}
\end{equation}
The difference $H_B^0-\widetilde H_B^0$ is invariant under the $i-$BT, since
\begin{equation}
    b_k^\dagger b_k -\tilde{b}^\dagger_k \tilde{b}_k = a^\dagger_k a_k -\tilde{a}^\dagger_k \tilde{a}_k\label{eq:fundamental invariance}
\end{equation}
holds for 2-mode Bogoliubov transformations with arbitrary $\theta_k$, and therefore also
for the inverse transformation.  This is a fundamental invariance which makes the Liouville gauge particularly convenient for real-time propagation: the thermofield bath Hamiltonian retains its original form, without mixed physical--tilde kinetic terms. 

For an anharmonic bath, Eqs.~(\ref{eq:Utheta}-\ref{eq:iBT}) remain a valid change of representation, but the transformed thermal state $\ket{\Omega^\theta(\beta)}=U_\theta^\dagger\ket{0(\beta)}$ is generally not the vacuum. Its residual occupations and correlations encode departures from the harmonic reference. 
%This differs from the generalized construction of Ref.~\cite{Gelin2017}, which uses a generator ansatz acting within the subspace spanned by matched physical--auxiliary energy eigenstates. The harmonic-reference unitary used here acts throughout the doubled space and is not tailored to map the anharmonic thermal purification onto the reference vacuum. 
Nevertheless, for the bosonic environments considered here, the transformed thermal state satisfies an identity similar to Eq.~\eqref{eq:thermal condition}
\begin{equation}
    \braket{\Omega^\theta(\beta)|A|\Omega^\theta(\beta)}=\braket{\Omega^\theta(\beta)|\tilde{A}^\dagger|\Omega^\theta(\beta)}\label{eq:thermal condition BT}
\end{equation}
This can be proved by noting that 
\[\braket{\Omega^\theta(\beta)|A|\Omega^\theta(\beta)}=\braket{0(\beta)|U_\theta A U_\theta^\dagger|0(\beta)}\] 
and using Eq.~\eqref{eq:thermal condition} with $A^{\bar{\theta}}=U_\theta A U_\theta^\dagger$ (the \emph{forward} Bogoliubov transformed operator $A$) in place of $A$. This is not obvious at first sight since the operator $A^{\bar{\theta}}$ is no longer physical-only (as required by Eq.~\eqref{eq:thermal condition}), due to the Bogoliubov mixing of physical and auxiliary variables. Nevertheless, for the bosonic case considered here \footnote{The troubles with fermions arise from a global fermionization of the double-space construct that is highly convenient for exploiting at full the canonical anti-commutation rules of fermion operators. That is, the appropriate space is the Fock space built out from the doubled single-particle space, rather than the space obtained by doubling the Fock space that originates from the physical single-particle space.}, physical and tilde operators commute with each other, and Eq.~\eqref{eq:thermal condition} generalizes to arbitrary operators in double space. This can be readily seen by considering first product operators of the type $A\otimes\tilde{B}$, i.e. proving
\[\braket{0(\beta)|A\otimes\tilde{B}|0(\beta)}=\braket{0(\beta)|B^\dagger\otimes\tilde{A}^\dagger|0(\beta)},\] 
and later extending the result to the sums of products. 
Therefore, given that $\tilde{U}_\theta\equiv U_\theta$ holds with the unitary  $U_\theta$ defined in Eq.~\eqref{eq:Utheta}, we have
\begin{equation}
    \widetilde{A^{\bar{\theta}}}^\dagger = \widetilde{\left(U_\theta A U_\theta^\dagger\right)}^\dagger = \left(U_\theta \tilde{A} U_\theta^\dagger\right)^\dagger=U_\theta \tilde{A}^\dagger U_\theta^\dagger 
\end{equation}
where $U_\theta^\dagger\ket{0(\beta)}\equiv\ket{\Omega^\theta(\beta)}$, thereby completing the proof of Eq.~\eqref{eq:thermal condition BT}.

We note that the harmonic-reference transformation used here differs from the generalized thermal transformation proposed in Ref.~\cite{Gelin2017}. The latter uses a generator ansatz acting within the subspace spanned by matched physical--auxiliary energy eigenstates. Our unitary acts throughout the doubled space and is not tailored to map the anharmonic thermal purification onto the reference vacuum.

\section{Thermalization in Bogoliubov picture}
\subsection{Imaginary-time dynamics}\label{sec:imaginary-time preparation}
Let $\ket{0_0(\beta)}$ be the thermal purification of the reference Hamiltonian $H_B^0$ and $Z_0$ the corresponding partition function, 
\begin{equation}
    \ket{0_0(\beta)}=\frac{1}{\sqrt{Z_0}}e^{-\beta H_B^0/2}\ket{I}=\frac{1}{\sqrt{Z_0}}e^{-\beta \tilde{H}_B^0/2}\ket{I}\label{eq:reference thermal state}
\end{equation}
Similarly, let 
\begin{equation}
    H_B = \sum_{k=1}^{N_B} \frac{p_k^2}{2m} + V(q_1,q_2,..q_{N_B})
\end{equation}
be the bath Hamiltonian of interest, $\ket{0(\beta)}$ the thermal state and $Z$ its partition function. Inverting the second identity of Eq.\eqref{eq:reference thermal state}, we can write $\ket{0(\beta)}$ as an imaginary-time evolved state from the reference 
\begin{equation}
 \ket{0(\beta)}=\frac{1}{\sqrt{Z}}e^{-\beta H_B/2}\ket{I}=\sqrt{\frac{Z_0}{Z}} e^{-\beta K_B/2}\ket{0_0(\beta)}
 \label{eq:referencefilter}
\end{equation}
with the thermalizing bath operator $K_B$ given by
\begin{equation}
    K_B = H_B-\widetilde H_B^0
\end{equation}
which correctly reduces to the Liouville-gauge thermofield generator when $H_B=H^0_B$ \footnote{Under these conditions, the thermal state is a zero-energy eigenstate of the generator, hence $\ket{0_0(\beta)}$ is stable against thermalization}. 
A reference-state imaginary-time conversion of this form was also discussed by Gelin and Borrelli~\cite{Gelin2017} for correlated thermal preparations. % Here, we examine its numerical conditioning when a harmonic reference is used to prepare an anharmonic target state.} 
%\st{For $H_B=H_B^0$ the thermalizing operator reduces to the Liouville-gauge thermofield generator of the reference, $H^0_B-\tilde{H}^0_B$, and the thermal state is a zero-energy eigenstate of this operator; thus, $\ket{0_0(\beta)}$ is stable against thermalization under these conditions, as it should be. }

Notably, upon monitoring the change of the norm of the imaginary-time evolving vector  $\ket{\Psi(\alpha)}$, one can compute the partition function $Z$ or the corresponding free-energy $F=-k_B T \ln Z$, according to
\begin{equation}
    F(\beta)=F_0(\beta)+2k_B T \int_0^{\beta/2} \bar{K}_B(\alpha)\,d\alpha
\end{equation}
where 
\begin{equation}
     \bar{K}_B(\alpha) = \frac{\braket{\Psi(\alpha)|K_B|\Psi(\alpha)}}{\braket{\Psi(\alpha)|\Psi(\alpha)}}
\end{equation}
is the expectation value of the thermalizing operator along the imaginary-time trajectory and $F_0(\beta)$ is the free-energy of the reference.

This is general, and holds for arbitrary reference Hamiltonians. For a harmonic reference we can perform the Bogoliubov transformation described in Sec. \ref{sec:thermal BT}. Applying $U_\theta^\dagger$ transforms Eq.~\eqref{eq:referencefilter} into a thermalization that starts from $\ket{\textbf{0}}$ and ends at $\ket{\Omega^\theta(\beta)}$, using the $i-$BT generator $K_B^\theta$,
\begin{equation}
    K_B^\theta = (H_B^0-\tilde{H}_B^0)^\theta + V_c^\theta \equiv H_B^0-\tilde{H}_B^0 + V_c^\theta\label{eq:BT thermalizer}
\end{equation}
where we have exploited the fundamental invariance of the Bogoliubov transformation, Eq.~\eqref{eq:fundamental invariance}. Here, we have introduced the correction potential
\begin{equation}
V_c=H_B-H_B^0:=V(q_1,q_2,..q_{N_B})-\sum_{k=1}^{N_B}\frac{m\omega_k^2}{2}q_k^2    
\end{equation}
and its Bogoliubov picture representation (see Eq.~\eqref{eq:BTV}), 
\begin{equation}
V_c^\theta=V(q^\theta_1,q^\theta_2,..q^\theta_{N_B})-\sum_{k=1}^{N_B}\frac{m\omega_k^2}{2}(q^\theta_k)^2    
\end{equation}
where each $q^\theta_k$ is the linear combination of $q_k$ and $\tilde{q_k}$ described in Eq.~\eqref{eq:BTq}.   

Although algebraically exact and physically appealing, the construction of Eq.~\eqref{eq:referencefilter} (and its Bogoliubov-picture variant) is poorly conditioned numerically. The factor $e^{+\beta\widetilde H_B^0/2}$ amplifies errors in the initially suppressed auxiliary components. Across an auxiliary energy span $\Delta E_0$, the relative amplification can scale as $e^{\beta\Delta E_0/2}$. Increasing the underlying basis can therefore make the thermalization less stable even while improving the representation of the desired state. For anharmonic finite-depth wells the reference subtraction is especially problematic because the reference energy continues to grow while the physical potential approaches its asymptote. 

Stated differently, while imaginary-time filtering with a physical Hamiltonian bounded from below suppresses high-energy components and provides a robust preparation route for the bare thermofield representation \cite{Venrell2026}, the reference-subtracted generator is \emph{not} generally bounded from below and easily becomes unstable under imaginary-time propagation. The auxiliary contribution $-\tilde{H}_B^0$ cannot simply be dropped from $K_B$ while retaining the specific conversion in Eq.~\eqref{eq:referencefilter} since, unlike real-time auxiliary rotations, an auxiliary imaginary-time factor is nonunitary and generally affects the physical state represented by the purification. 

The observed symptoms of instability include discrepancies between physical and auxiliary moments and eventually large reference number occupations in the Bogoliubov picture. Tighter integrator tolerances reduce avoidable errors but cannot remove this conditioning problem. Agreement of a few low-order moments on a small grid is also insufficient when the subsequent coupled system-bath dynamics requires higher excitations. These considerations motivated the direct constructions of the thermal state described in the next Sections.

\subsection{Direct pair construction}\label{sec:direct pair-construction}
We start with a one-step, direct pair construction (DC) that applies to independent oscillator bath models. Denoting $h_k$ the $k$-mode Hamiltonian, let us consider its thermal amplitude
\begin{equation}
 A_k=\frac{e^{-\beta h_k/2}}{\sqrt{\textrm{Tr}\left(e^{-\beta h_k}\right)}},\qquad
 \ket{0_k(\beta)}=\sum_{ij}(A_k)_{ij}\ket{i}\ket{\widetilde j}
 \label{eq:amplitude}
\end{equation}
This can be obtained upon exact or iterative diagonalization.  
Application of $U_{\theta,k}^\dagger$ to the resulting double-space vector would give us the desired amplitude in the Bogoliubov picture for the $k$-mode pair. 

Let us then consider the finite-basis version of $U_{\theta,k}^\dagger$ obtained by representing $a_k$ and $\tilde{a}_k$ in a finite basis. We conveniently chose the latter to be a truncated HO eigenbasis of the same mode frequency but this is not mandatory, provided the matrix elements of the annihilation operators are available. The generator in Eq.~\eqref{eq:Utheta} is anti-Hermitian and its exponential is therefore unitary in that finite space. This guarantees norm preservation for the discrete transformation, although agreement with the infinite-dimensional transformation has to be checked. To initialize a pair in a discrete variable representation (the HO-DVR for the choice above), we reshape the coefficients of the transformed thermal amplitude into a matrix and compute its singular value decomposition, which amounts to writing 
\begin{equation}
O^\theta(q,\tilde{q})=\braket{q,\tilde{q}|U_{\theta,k}^\dagger|0_k(\beta)}=\sum_l s_l u_l(q) v_l(\tilde{q})    
\end{equation}
This provides a sum of products that can act as local operators on the uniform vectors $\phi_1(q)=\phi_2(\tilde{q})\equiv 1$ to initialize the MCTDH wavefunction with the correct $k-$mode thermal wavefunction,
\begin{equation}
    \Omega^\theta_k(\beta)(q,\tilde{q})=O^\theta(q,\tilde{q})\phi_1(q)\phi_2(\tilde{q})
\end{equation}
Independent pair states can be initialized directly in an ML-MCTDH wavefunction \cite{mctdh:MLpackage}; the only caveat is that operator applications need to be staged in groups of at most fifteen to respect the package limit. 

If desired, the truncated mode Hamiltonian can be modified, prior to construction, to project out undesired components. For instance, with dissociative modes one can discard the continuum eigenstates and only retain the purely bound component of the Hamiltonian. 

\subsection{Transformation after bare thermalization}\label{sec:direct route to BT}
A second route, denoted "apply-after-thermalizing" (AA), retains the generality of imaginary-time propagation to overcome the limitations of an independent oscillator model. We start from the identity vector to prepare the bare thermal state and later apply the (discretized) Bogoliubov unitary introduced above,
\begin{equation}
 \ket{0(\beta)}\propto e^{-\beta H_B/2}\ket{I},\,\,\,
 \ket{\Omega^\theta(\beta)}=\left(\prod_{k=1}^{N_B}U_{\theta,k}^\dagger\right)
 \ket{0(\beta)}.
 \label{eq:after}
\end{equation}
An equivalent symmetric filter $e^{-\beta(H_B+\widetilde H_B)/4}$ can be used to enforce the physical-auxiliary symmetry of the thermal state, but this is not required as this thermalization is stable and displays the expected symmetry to numerical precision. 

Bare filtering avoids the inverse reference factor, at the price of introducing a discretized version of the Bogoliubov transformation.  For ML-MCTDH, a physical--tilde pair is placed in a combined mode, and the mode-operator $U_{\theta,k}^\dagger$ is applied to the bare-thermalized wavefunction. To this end, the DVR matrices of the transformation operators 
%\[V_{\alpha,\alpha'|\beta\beta'}=\braket{q^\alpha_k\tilde{q}^{\alpha'}_k|U_{\theta,k}^\dagger|q^\beta_k\tilde{q}^{\beta'}_k}\]
\[V_k(q,\tilde{q}|q',\tilde{q}')=\braket{q,\tilde{q}|U_{\theta,k}^\dagger|q'\tilde{q}'}\]
are directly read in MCTDH \cite{mctdh:MLpackage} and, as above, their application is staged in groups of at most fifteen to respect the package limit. 

The advantage of this approach is that Equation~\eqref{eq:after} remains valid when $H_B$ contains interactions between bath modes: the bare thermalization then creates intermode correlations before local pair transformations are applied. The disadvantage is that modifications such as the bound-state projection mentioned above need to be addressed in the thermalization step. %The automatic implementation used here assumes the original mode Hamiltonians.

\subsection{Numerical analysis}
\begin{figure*}
\centering
\includegraphics[width=0.8\textwidth]{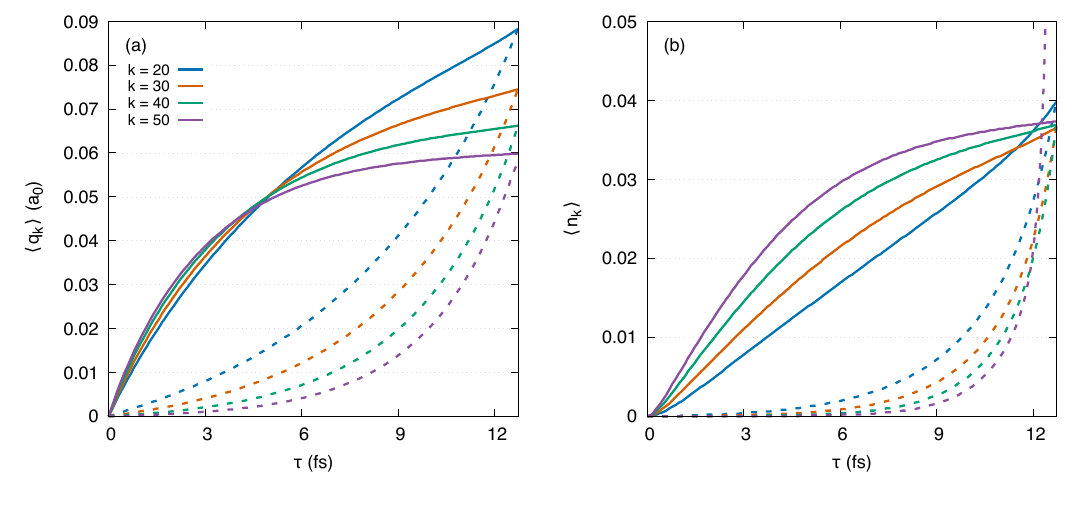}
\caption{Bogoliubov-picture imaginary-time preparation at $T=300$ K for a Morse oscillator bath with $\xi=0.100$. Representative imaginary-time trajectories for modes $k=20,30,40,50$, all using $N=8$ HO-DVR primitive functions per coordinate. Colors identify the modes, solid and dashed lines identify physical and auxiliary coordinates. Coordinate (panel a) and occupation-number (panel b) expectation values taken with the Bogoliubov-picture wavefunction. The endpoint is $\tau_\beta=12.7305$ fs. Panel (b) is restricted to $\langle n_k\rangle\leq0.05$; the auxiliary $k=50$ endpoint ($0.1792$) lies outside the displayed range.}\label{fig:paths}
\end{figure*}

The numerical instability of the Bogoliubov-picture imaginary-time propagation is illustrated at $T=300$~K for a Morse oscillator bath at the moderate anharmonicity value of $\xi=0.100$ (see Sec. \ref{sec:MO bath} for a description of the bath model). The initial ground-state wavefunction of the reference HO Hamiltonian is propagated with the thermalizing operator $K_B^\theta$ of Eq.~\eqref{eq:BT thermalizer} up to $\tau_\beta = \beta\hbar/2 = 12.7305$ fs, using the accurate VMF integrator \cite{mctdh:MLpackage} with ABM/all = 6, $10^{-10}$, $10^{-4}$. 

Figure~\ref{fig:paths} shows coordinate and occupation-number expectations taken with the Bogoliubov-picture wavefunction, for four physical-auxiliary pairs, using eight primitive functions per coordinate (HO-DVR). Since the thermalizing generator acts asymmetrically on the two members of a pair, unequal physical and auxiliary paths at intermediate imaginary times are not themselves evidence of error. The relevant test is the state reached at the thermal endpoint. For $k=50$, the endpoint occupation-number expectations are $0.0373$ and $0.1792$, despite the much smaller discrepancy between the coordinate means, $0.0598$ and $0.0580\,a_0$. Number expectations thus expose a defect that is less apparent in the first coordinate moment.

\begin{figure*}
\centering
\includegraphics[width=0.7\textwidth]{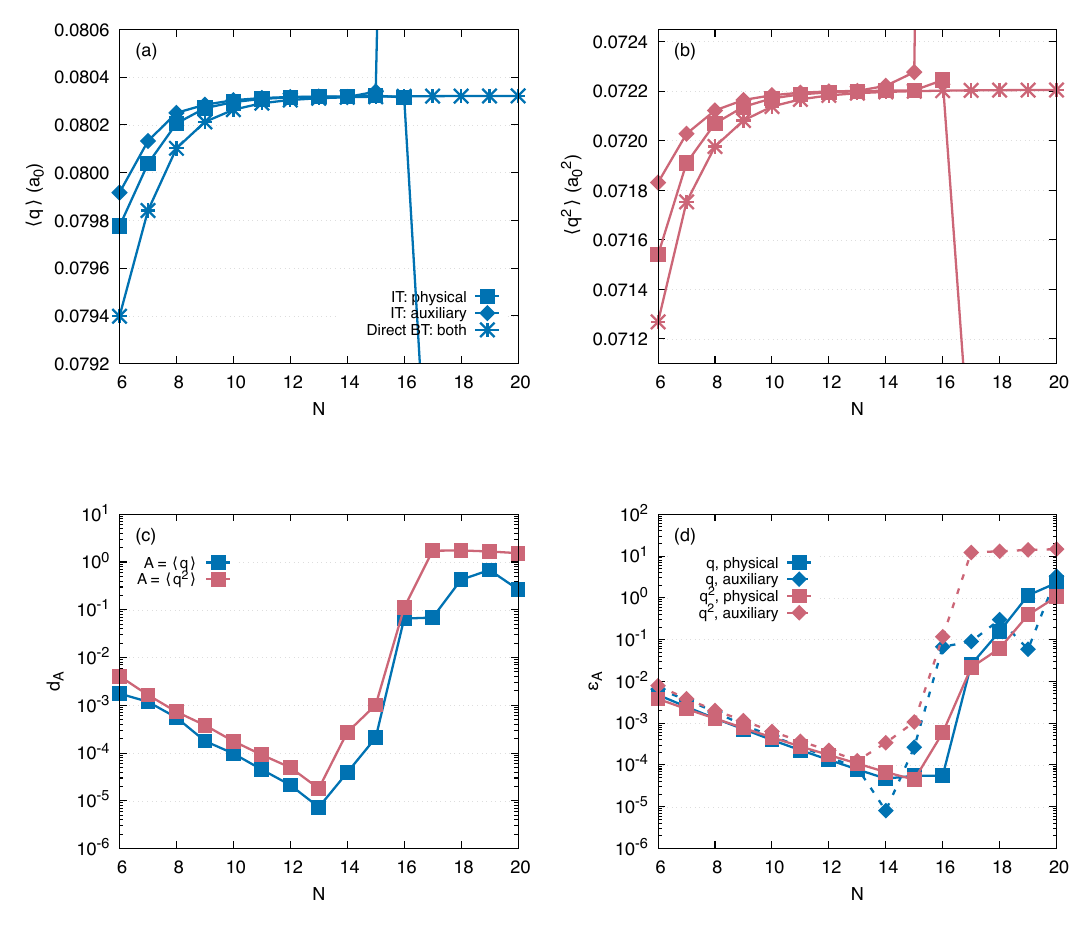}
\caption{Primitive-basis stability for the $(q_{25},\tilde{q}_{25})$ pair, at $T=300$ K for the same Morse oscillator bath considered in Fig.\ref{fig:paths} ($\xi=0.100$). (a,b) Mean and second moment after imaginary-time propagation (IT), compared with apply-after-thermalizing (AA scheme for direct BT, see main text for details). The upper panels zoom into the convergence region; unstable values outside these ranges are omitted there but are included in the lower diagnostics. (c) Relative physical--auxiliary symmetry defects, $d_A=|A-\widetilde A|/|(A+\widetilde A)/2|$, where $A$ denotes the indicated expectation value. (d) Relative differences from direct BT at the same $N$, $\epsilon_A=|A_{\mathrm{IT}}-A_{\mathrm{BT}}|/|A_{\mathrm{BT}}|$.}\label{fig:basis}
\end{figure*}

A primitive-basis scan of the $q_{25}$--$\widetilde q_{25}$ pair provides a more stringent demonstration (Fig.~\ref{fig:basis}). Let us define the relative symmetry defect as
\begin{equation}
 d_A(N)=\frac{|\langle A\rangle_N-\langle\widetilde A\rangle_N|}
 {|[\langle A\rangle_N+\langle\widetilde A\rangle_N]/2|}
\end{equation}
for $A=q,q^2$, with $N$ denoting the primitive-basis size. Both defects decrease initially and attain their smallest sampled values at $N=13$: $d_q=7.34\times10^{-6}$ and $d_{q^2}=1.81\times10^{-5}$. Increasing the basis further reverses this trend. At $N=16$, the defects have grown to approximately $6.54\%$ and $11.04\%$, respectively. At $N=20$, the second moments are $0.148889$ and $1.127065\,a_0^2$, whereas apply-after-thermalizing gives $0.0722050\,a_0^2$ for both pair members. The corresponding auxiliary occupation number expectation is $18.97$, close to the highest reference oscillator index, $N-1=19$. This is consistent with amplification of high-lying auxiliary components.

The apply-after-thermalizing moments remain paired to all reported digits and continue to settle as the basis increases. Between $N=18$ and $20$, their relative changes are $6.93\times10^{-6}$ for the mean and $1.20\times10^{-5}$ for the second moment, suggesting a good degree of finite-basis convergence. Figure~\ref{fig:basis}(d) separately compares the two preparation routes, showing that finite-basis operations can contribute to their differences for $N \lesssim 13$, while instability in the reference-thermalization route dominates at larger $N$, where the apply-after-thermalizing moments show stable finite-basis convergence. 

For the independent-oscillator bath considered here, we also checked direct pair construction against apply-after-thermalizing using the same finite-basis Hamiltonians and transformation matrices. The two routes gave numerically indistinguishable prepared states (data not shown). %This agreement checks the two implementations within the chosen discretization; convergence of the primitive basis remains a separate requirement. 

\section{Morse Oscillator bath models}\label{sec:MO bath}
\subsection{Generalized Caldeira--Leggett Hamiltonian}
The environment extends the displaced-oscillator structure of the Caldeira--Leggett model \cite{Caldeira1983,Weiss2008} by replacing the harmonic coordinate with a nonlinear displacement \cite{MorseBath2011},
\begin{align}
 H&=H_S+\sum_{k=1}^{N_B}\left[\frac{p_k^2}{2M}
 +\frac{M\omega_k^2}{2}\left(g_{\alpha_k}(q_k)
 -\frac{c_k f(s)}{M\omega_k^2}\right)^2\right]\label{eq:generalized CL model}
 \end{align}
 where
\begin{equation}
 g_\alpha(q)=\frac{1-e^{-\alpha q}}{\alpha},\qquad
 \alpha_k=\sqrt{\frac{M\omega_k\xi}{\hbar}}   
\end{equation}
and $f(s)$ is a function shaping the interaction with the subsystem coordinate $s$ ($f(s)= s$ for a coupling linear in $s$). The harmonic limit follows as $\xi\to0$, since $g_\alpha(q)\rightarrow q$ under this condition. Expanding the square identifies the interaction 
\begin{equation}
H_I=-f(s)\sum_{k=1}^{N_B}c_kg_{\alpha_k}(q_k)\label{eq:H_I}     
\end{equation}
and the counterterm
\begin{equation}
\Delta v(s)=f(s)^2\sum_{k=1}^{N_B} \frac{c_k^2}{2M\omega_k^2}\label{eq:counterterm}.    
\end{equation}

The uncoupled Morse well has depth and bound-state energies
\begin{align}
 D_k&=\frac{\hbar\omega_k}{2\xi},\\
 E_{k.n}&=\hbar\omega_k\left(n+\frac12\right)
 -\frac{\hbar\omega_k\xi}{2}\left(n+\frac12\right)^2
 \end{align}
 where
\[
 n+\frac12<\frac1\xi.\label{eq:morsespectrum}
\]
(here the strict inequality excludes a threshold state). This shows that $\xi$ -- the single model parameter governing the bath anharmonicity -- has the simple, approximate meaning of inverse number of bound states present in each oscillator of the bath. For instance, at $\xi=0.1$ and $0.5$, each oscillator of the bath supports, respectively, ten and two bound levels. Furthermore, upon computing the adjacent bound-state transition energy 
\begin{equation}
\Delta E_{k,n}=E_{k,n+1}-E_{k,n}=\hbar\omega_k[1-\xi(n+1)]\label{eq:adjacent De}
\end{equation}
it becomes clear that energy uptake shifts a mode away from its reference resonance. The bath therefore acquires occupation-dependent transition frequencies and a finite bound-state excitation capacity. 

In the calculations described below the reference frequencies and couplings were drawn from a truncated Ohmic spectral density 
\[J(\omega)=m\gamma\omega\Theta(\omega_c-\omega)\] 
where $\gamma$ is the model relaxation rate, $m$ is the system mass and $\omega_c$ the frequency cutoff, set to $2000\,\mathrm{cm}^{-1}$, twice the subsystem frequency in the high-frequency model. We use the conventional definition of spectral density for the Caldeira-Leggett Hamiltonian, 
\begin{equation}
J(\omega)=\frac{\pi}{2}\sum_{k=1}^{N_B} \frac{c_k^2}{M_k\omega_k}\delta(\omega-\omega_k) 
\end{equation}
noticing that for an anharmonic bath the latter specifies the parametrization (the bath frequencies $\omega_k$ and the couplings $c_k$) 
but no longer determines the thermal force correlations and the memory kernel. 

Unless otherwise stated, the calculations used $N_B=50$ physical modes up to $\omega_c=2000\,\mathrm{cm}^{-1}$ and explored $0\leq\xi\leq0.5$. The nominal harmonic recurrence scale $t_{\textrm{rec}}=2\pi/\Delta\omega=2\pi N_B/\omega_c$ is about $0.8$ ps, although this remains a rough estimate, as anharmonic shifts modify also recurrence behavior. The temperature was varied in the range $T=0-400\,K$ but the systematic results presented here refer  to $T=300\,K$. The finite-temperature vibrational-relaxation series uses full bath doubling, giving $100+1$ coordinates. 

Beyond the present model calculations, as observed above, substantial savings are possible by noting that in thermal equilibrium high-frequency modes ($\hbar\omega_k\gg k_B T$) are essentially in their ground state and need not be doubled when the bath is thermalized in the initial state. We used this strategy in some decoherence calculations, where thermally inactive auxiliaries were omitted, as documented in the Supplementary Material. 

\subsection{Thermal preparation of dissociative modes}
Since the Morse potential is dissociative, 
its unrestricted, finite-temperature canonical partition function on an infinite line diverges. A finite-grid Gibbs state inevitably includes box-dependent continuum contributions. Accordingly, the initial condition can be specified as a confined thermal state, a bound-conditioned state, or a prepared metastable state, and these are physically distinct choices.

For independent modes, an explicit bound-conditioned ensemble can be constructed with the projector onto the bound-state component $P_b$ 
\begin{equation}
 \rho_{B,b}=\frac{P_b e^{-\beta H_B}P_b}{\textrm{Tr}(P_be^{-\beta H_B})}.
\end{equation}
Such a projector restricts the initial state, without constraining the Hilbert space used for the coupled dynamics. Dissociation induced subsequently by subsystem coupling can still occur if the primitive functions support the continuum. The correct description of the dynamics then requires the introduction of complex absorbing potentials (CAPs). These alleviate the unphysical reflection from the grid boundaries that would alter the dynamics and would make integration of the equation-of-motion numerically stiff. This issue is most critical for the low-frequency modes that present dissociation energies comparable to the thermal energy, already for small values of the anharmonicity. 

In our protocol, low-frequency modes are kept in the bare thermofield representation on extended FFT-DVR grids. After finite-box thermalization, prolonged uncoupled real-time propagation with CAPs along both physical and auxiliary modes removes much of the extended component, followed by normalization. 
This CAP-filtering strategy also applies to interacting baths, provided their internal couplings are retained. It produces an approximately bound initial state similar to the spectral projection above, at the cost of an additional real-time propagation involving only the bath modes.

A CAP placed beyond the bound-state region minimizes perturbation of localized components, but near-threshold components escape slowly and bound-state tails can also be absorbed. Preparation time, grid coverage, and CAP parameters need to be tested. In our protocol potentially dissociative grids are selected empirically using $D_{0,k}\lesssim3k_B\max(T,T_{\rm grid})$, where $D_{0,k}=D_k-E_{k,0}$ is the dissociation energy (corrected for the zero-point-energy) and $k_BT_{\rm grid}$ is an empirical energy scale chosen to accommodate bath excitation during the coupled dynamics (see Supplementary Material). Furthermore, an uncoupled propagation of $10$ ps was found sufficient to absorb the unbound component and was used throughout \footnote{For this bath-only propagation, our independent-oscillator model requires only a time-dependent self-consistent-field product wavefunction over bath modes, with each physical--auxiliary pair treated as a single combined mode.}.

Coupled dynamics used CAPs along physical modes only. For these untransformed low-frequency modes, the bare-gauge generator contains no auxiliary contribution, so absorption is applied only to physical outgoing components. In this way, the norm loss during the coupled dynamics becomes a measure of the MO-bath dissociation due to energy exchange with the subsystem. We remark that, in the presence of CAPs, all ordinary expectation values are normalized by the instantaneous squared wavefunction norm and therefore refer to the surviving component.

\subsection{Bogoliubov modes}

In our protocol, high-frequency modes are Bogoliubov transformed as described in Sec. \ref{sec:direct route to BT}. Their coupled dynamics within the generalized Caldeira-Leggett Hamiltonian of Eq.~\eqref{eq:generalized CL model} requires that the latter undergo the $i-$BT for each mode treated in this way. As discussed in Sec.~\ref{sec:thermal BT}, subtracting the auxiliary harmonic-reference Hamiltonian allows one to exploit the invariance in Eq.~\eqref{eq:fundamental invariance}. For the anharmonic bath, this differs from subtracting the full auxiliary Hamiltonian in the Liouville gauge. For the Morse-bath Hamiltonian of Eq.~\eqref{eq:generalized CL model} we choose the $k$-mode generator of the real-time dynamics to be $h_k-\tilde{h}^0_k$, so that its Bogoliubov form reads as
\begin{equation}
    (h_k-\tilde{h}^0_k)^\theta =h^0_k-\tilde{h}^0_k + v_{c,k}^\theta
\end{equation}
where 
\begin{equation}
    v_{c,k}^\theta = \frac{M\omega_k^2}{2}\left(g^2_{\alpha_k}(q^\theta_k)-(q^\theta_k)^2\right) 
\end{equation}
is the $i-$BT anharmonic correction.
This term can be written in sum-of-product form upon noticing that (with $(\mathcal{C}_k,\mathcal{S}_k)=(\cosh,\sinh)(\theta_k)$)
\begin{align}
    \alpha_k^2 g^2_{\alpha_k}(q^\theta_k)&=1-2 e^{-\alpha_k \mathcal{C}_k q_k}e^{-\alpha_k \mathcal{S}_k \tilde{q}_k}\\
    & + e^{-2\alpha_k \mathcal{C}_k q_k}e^{-2\alpha_k \mathcal{S}_k \tilde{q}_k}
\end{align}
while, of course, 
\begin{align}
 (q^\theta_k)^2 = \mathcal{C}_k^2q^2_k+\mathcal{S}_k^2\tilde{q}^2_k+2\mathcal{C}_k\mathcal{S}_kq_k\tilde{q}_k
\end{align}
and
\begin{equation}
    h^0_k-\tilde{h}^0_k = \frac{p_k^2-\tilde{p}^2_k}{2M}+\frac{M\omega^2_k(q^2_k-\tilde{q}^2_k)}{2}
\end{equation}
The interaction term of Eq.\eqref{eq:H_I} undergoes the same transformation, and the $k-$mode contribution reads as
\begin{align}
H^\theta_{I,k}=&-c_k f(s)g_{\alpha_k}(q^\theta_k)\nonumber\\
=&-\frac{c_kf(s)}{\alpha_k}\left(1-e^{-\alpha_k \mathcal{C}_k q_k}e^{-\alpha_k \mathcal{S}_k \tilde{q}_k}\right)        
\end{align}
This completes the specification of the thermofield Hamiltonian used in the real-time dynamics of the coupled system, which we write here for convenience,
\begin{align}
    H_\textrm{TFD} = & H_S + \Delta v(s) +\nonumber \\
    & + \sum_{k\notin \mathcal{B}} h_k -f(s) \sum_{k\notin \mathcal{B}} c_k g_{\alpha_k}(q_k )+ \nonumber\\
    & + \sum_{k\in \mathcal{B} }\left(h^0_k-\tilde{h}^0_k + v_{c,k}^\theta\right) -f(s) \sum_{k\in \mathcal{B}} c_k g_{\alpha_k}(q^\theta_k)
\end{align}
where $\mathcal{B}$ denotes the set of high-frequency, Bogoliubov transformed modes.

\subsection{Wavefunction propagation and convergence}
ML-MCTDH represents the wavefunction through a hierarchy of time-dependent single-particle functions \cite{Wang2003,ven11:044135}. In our calculations at finite temperature, each retained physical--auxiliary pair was placed in a combined mode at the primitive basis level. Pair grouping is natural for thermofield initialization, as it allows one to accurately describe the thermal correlations between physical and auxiliary modes that are present in the thermal state of the bare thermofield description. It is also very useful under the Bogoliubov transformation, as the latter introduces coupling between physical and auxiliary modes, that can thus be accurately addressed at the primitive basis level. The remaining tree structure controls the representation of system--bath and bath--bath correlations, and was thoroughly checked for convergence and performance. Analysis of the natural populations identifies nodes that may need enlargement, and was systematically used to improve convergence, although small least-occupied natural populations do not guarantee overall convergence of the ML-MCTDH wavefunction (which remains tied to the choice of tree structure). 

The primitive basis was selected according to the mode type. As mentioned above, we distinguish two sets of bath modes: low-frequency modes that are not Bogoliubov transformed and that are treated with FFT-DVR grids and CAPs, and high-frequency modes that are Bogoliubov transformed and that are described with HO-DVR. As for the subsystem mode $s$ we opted for a dense FFT-DVR grid. The primitive basis was checked by monitoring selected observables, typically the average and the root mean squares of the position and of the occupation number operators for the bath oscillators. Several additional operators were used for the subsystem, including the system and the interaction energy. 

For the zero-temperature reference calculations, no Bogoliubov transformation is needed. Dissociative modes used uniform grids with physical CAPs, while the remaining modes used HO-DVR. In this case, an effective temperature $T_{\rm grid}=300$ K was set to retain the same dissociative-mode selection and physical CAP onsets and strengths as the finite-temperature series. The adopted grids were thus very similar to the finite-temperature ones, except for the odd point counts inherited from exponential-DVR initialization (although always followed by FFT propagation).
%The zero-temperature uniform grids use a wider left-boundary rule than the archived finite-temperature grids (coefficients 50 and 10, respectively), together with odd point counts inherited from exponential-DVR initialization and retained in FFT propagation. Minimum grid dimensions and a few low-anharmonicity representation choices also differ, as detailed in the Supplementary Material.
The Supplementary Material describes the primitive-grid rules, the CAP parameters and the tree structures adopted. 

Integration of the ML-MCTDH equation of motion used constant-mean-field (CMF) or variable-mean-field (VMF) propagation, with a preference for the split CMF algorithm. VMF propagation also served as reference for testing the less expensive CMF settings. It turned out that reliable real-time propagation in the investigated anharmonic environments requires both an accurately prepared thermal state and adequate integration tolerances. We further found that tightening the outer CMF accuracy to $10^{-8}$ is more effective than tightening a single local propagator. In this split-CMF setting, each mode propagator used RK8 with $10^{-8}$ accuracy; the exception is the top node that used SIL, with the same accuracy threshold and a maximal order of 20. Settings for the unified-CMF and VMF retained in some calculations, are specified in the Supplementary Material. Bath mode dynamics was typically monitored for convergence, since the CAPs make the propagation nonconservative even at small anharmonicity due to the introduction of CAPs.

\section{Vibrational relaxation}\label{sec:vibrational relaxation} 
\begin{figure*}[t]
\centering
\includegraphics[width=0.8\textwidth]{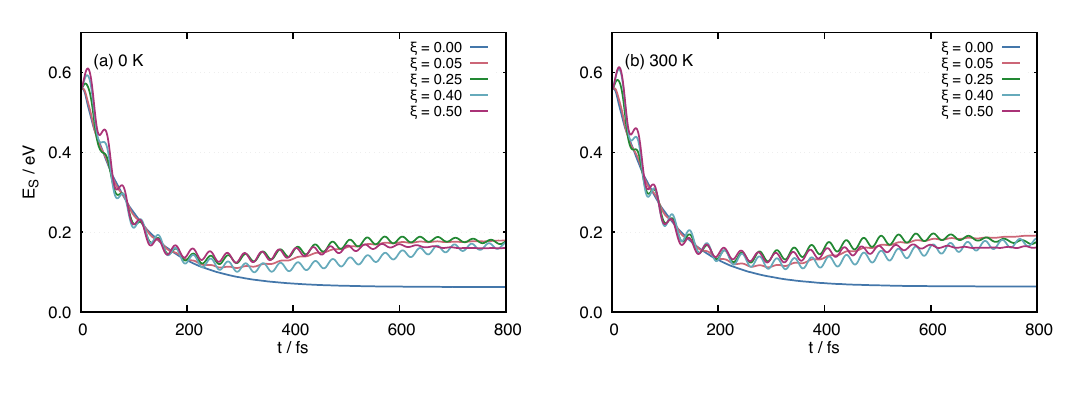}
\caption{Bare subsystem energy $E_S(t)$ following preparation in $v=4$ vibrational level, at (a) $0$ K and (b) $300$ K. The harmonic reference relaxes smoothly, whereas anharmonic baths produce pronounced oscillatory structure.}
\label{fig:relaxation-bare-energy}
\end{figure*}
\begin{figure*}[t]
\centering
\includegraphics[width=0.9\textwidth]{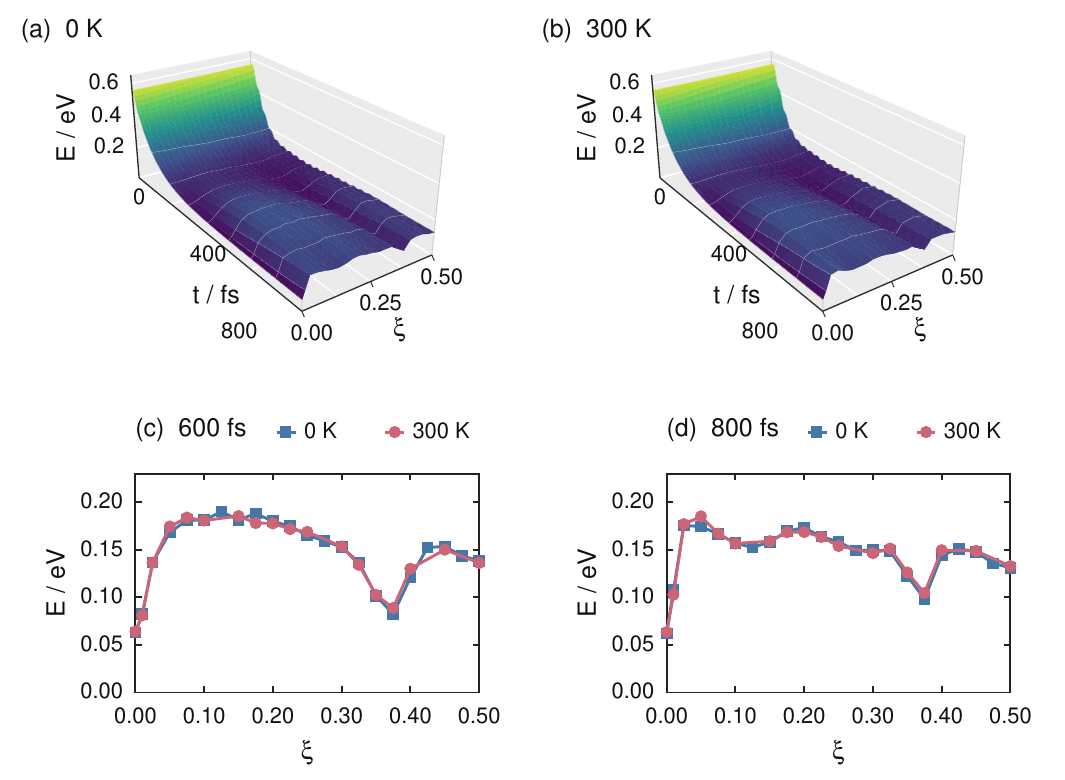}
\caption{Interaction-corrected subsystem energy $E_S^\textrm{int}=E_S+E_I/2$ following preparation in the $v=4$ vibrational level, for the bath at (a) $0$ K and (b) $300$ K. Long-time cuts at (c) $600$ fs and (d) $800$ fs compare the two temperatures; symbols mark every sampled $\xi$. The latter cut lies near the nominal harmonic-bath recurrence time.}
\label{fig:relaxation-energy}
\end{figure*}

We investigated vibrational relaxation of a harmonic oscillator with frequency $\omega_0=1000\,\mathrm{cm}^{-1}$ and the mass of a hydrogen atom. The simple choice of a harmonic oscillator was made to identify the effects of the \emph{bath} anharmonicity, without the complications of the system anharmonicity. For the same reason we used a state-independent friction model ($f(s)=s$). 

Two different initial states were considered, mimicking two different kinds of preparation experiments. In the first, the system was prepared in a high-lying vibrational eigenstate $\phi_v(s)$, as typically occurs after a resonant absorption process. In the second, the system was started as a localized wavepacket displaced from its equilibrium position, as it often happens when photo-exciting a ground-state species with a short-pulse laser on an electronic excited state, which is typically displaced with respect to the ground-state. In both cases the bath was started from its (uncoupled) thermal state and system-bath coupling was given the moderate-to-strong value of $\gamma^{-1}=100$ fs, in order to make stringent tests for the coupled dynamics and, at the same time, to observe a good degree of relaxation within the time-window set by the bath dimension (as mentioned above, $t_\textrm{rec}\approx 800$ fs with the current setting).
\subsection{Eigenstate relaxation}
The system was initially prepared in the high-lying $v=4$ vibrational state, $\psi_{S}(s)=\phi_4(s)$, and coupled to the independently prepared bath at $300$ K. This preparation carries a definite bare subsystem energy while placing substantial excitation into a bath whose bound-state capacity decreases as $\xi$ increases. The $T=0$ K case was also considered for comparison. We monitored the bare subsystem energy 
\[E_S(t)=\textrm{Tr}[H_S\rho_S(t)]\] 
and eigenstate populations 
\[P_v(t)=\bra{v}\rho_S(t)\ket{v}\]
as well as the average system coordinate ($\braket{s}$) and its spread ($\braket{ds}$),
and the subsystem purity
\[\mathcal{P}(t)=\textrm{Tr}(\rho_S(t)^2)\]
All these quantities are available as simple expectation values of the thermofield wavefunction except for purity, which requires the explicit calculation of the subsystem density operator $\rho_S(t)$. Purity starts from 1 with our preparation protocol and decreases as the subsystem becomes correlated with the bath, and may approach a thermal value if equilibration occurs. A useful reference value is provided by the uncoupled oscillator in thermal equilibrium, whose purity is
\[\mathcal{P}_\infty=\tanh\left(\frac{\beta\hbar\omega_0}{2}\right)\approx 0.984\,\,\textrm{at}\,\,T=300\,K\]
although the coupled subsystem is not expected to reach equilibrium within the simulated interval. 

To include system--bath coupling in the energy budget, we also
considered the symmetrically partitioned energy \cite{nest2003}
\begin{equation}
 E_S^{\rm int}(t)=E_S(t)+\tfrac12 E_I(t)\label{eq:shared-energy}
\end{equation}
where $E_I(t)$ is the expectation value of the interaction term of Eq.~\eqref{eq:H_I}.
This assignment provides a complementary diagnostic of energy
exchange, although the partition of interaction energy is not unique.

\begin{figure*}[t]
\centering
\includegraphics[width=\textwidth]{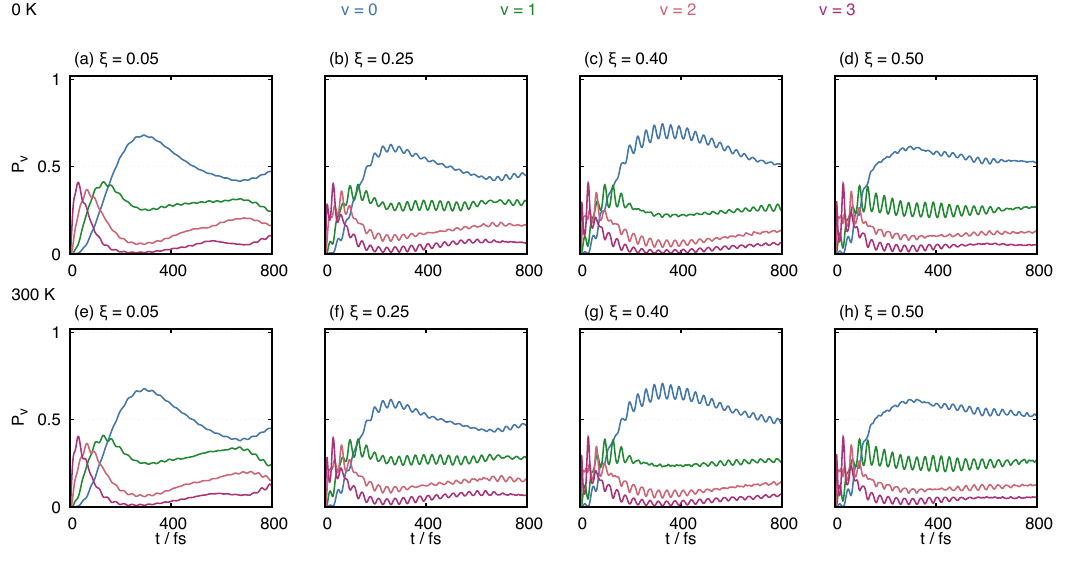}
\caption{Selected subsystem populations $P_v$ for $v=0,1,2,3$, at $0$ K (top row) and $300$ K (bottom row). Columns show $\xi=0.05$, $0.25$, $0.40$, and $0.50$, as indicated. The initial growth and subsequent decrease of $P_0$ accompany renewed excited-state populations.}
\label{fig:relaxation-populations}
\end{figure*}

The harmonic reference displays rapid energy loss followed by approach to a low-energy state over the available time window [see the $\xi=0.00$ curves in Fig.~\ref{fig:relaxation-bare-energy}(a,b)]. Its weak temperature dependence is consistent with $\hbar\omega_0\gg k_BT$ at $300$ K. Anharmonic baths alter the later dynamics markedly: energy decay develops oscillations and substantial residual excitation. The oscillations are reduced when half of the interaction energy is included in $E_S^{\rm int}$ [Eq.~\eqref{eq:shared-energy}], indicating that reversible exchange with the interaction term contributes appreciably to the bare-energy modulation. Nevertheless, oscillations remain visible in the shared energy, especially at larger $\xi$ (Fig.~\ref{fig:relaxation-energy}).

The late-time cuts in Fig.~\ref{fig:relaxation-energy}(c,d) show a strongly non-monotonic dependence on $\xi$, with closely similar structure at $0$ and $300$ K. Both temperatures display a local region of more efficient relaxation near $\xi=0.375$, visible at both $600$ and $800$ fs. This minimum remains below the $\xi=0.50$ endpoint at both times and temperatures. At the common sampled anharmonicities, the root-mean-square difference between the two temperature cuts is approximately $4$ meV at both times. Temperature modifies the residual energy and the oscillatory dynamics, but these changes are small compared with the non-monotonic variation induced by anharmonicity and do not substantially relocate the minimum. The near-resonant bath transitions lie at frequencies large compared with $k_BT/\hbar$, so their weak initial thermal population provides a physical reason for this similarity; it does not exclude thermal effects mediated by lower-frequency modes (including differences in their dissociation probabilities). Comparing the two cuts also shows the remaining time dependence: their common structure should not be interpreted as a stationary equilibrium curve or as evidence that recurrence effects are absent. The population dynamics helps characterize the retained excitation (Fig.~\ref{fig:relaxation-populations}). At intermediate anharmonicities, $P_0$ initially rises and then decreases as the excited-state populations grow again, particularly those of $v=2$ and $3$. These marginal populations reveal renewed subsystem excitation, without identifying a unique transition pathway. At large anharmonicity, appreciable excited-state populations persist, with oscillations about slowly varying mean values rather than clear equilibration over the observed interval.

\begin{figure*}[t]
\centering
\includegraphics[width=\textwidth]{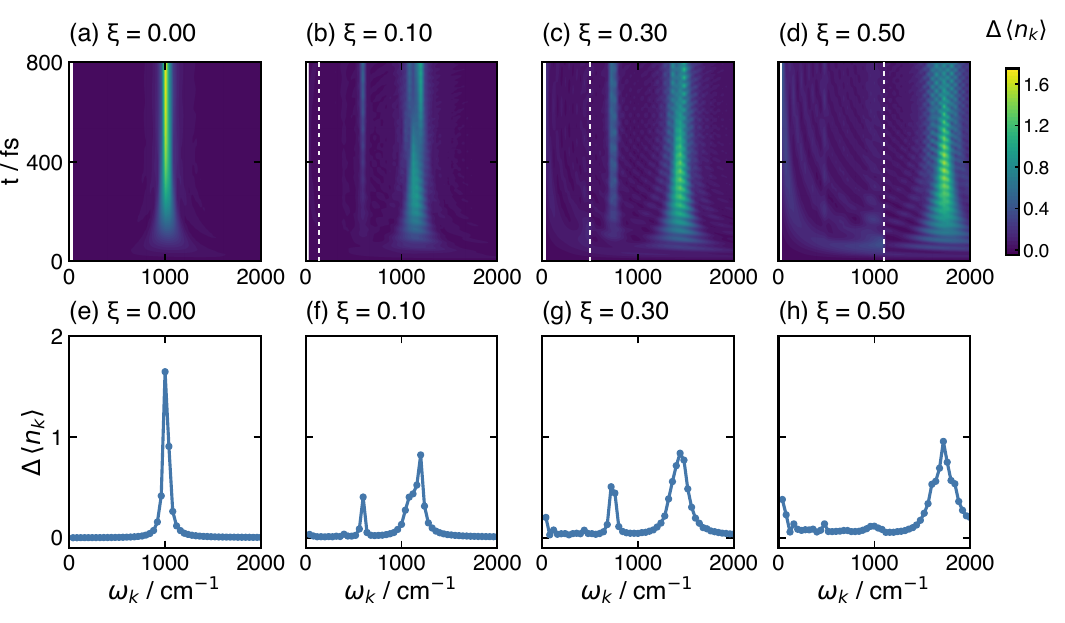}
\caption{Zero-temperature bath response for $\xi=0.00$, $0.10$, $0.30$, and $0.50$ (columns). Top: changes $\Delta\langle n_k\rangle=\langle n_k(t)\rangle-\langle n_k(0)\rangle$ in harmonic-reference occupations, on a common color scale; filled contours interpolate between sampled modes. White dashed lines mark the change from FFT to HO-DVR, halfway between adjacent mode frequencies (140, 500, and $1100\,\mathrm{cm}^{-1}$ in the anharmonic panels, respectively); the harmonic reference uses HO-DVR throughout. Bottom: the same occupation changes averaged over $500$--$750$ fs. The reference frequencies are spaced by $\Delta\omega=40\,\mathrm{cm}^{-1}$. The smooth response across the marked boundaries provides a consistency check between primitive representations. Comparison with the harmonic reference highlights the shift and broadening of the main resonance and the appearance of additional peaks.}
\label{fig:relaxation-modes}
\end{figure*}

The Morse spectrum suggests two competing effects: anharmonicity changes which bath transitions can accept a subsystem quantum, while excitation of a resonant mode changes its subsequent transition frequency. For the $n=0\rightarrow1$ transition, Eq.~\eqref{eq:adjacent De} predicts the resonant reference frequency
\[\omega_{\rm res}=\frac{\omega_0}{1-\xi}.\]
The narrow harmonic-bath response near $\omega_0$ in Fig.~\ref{fig:relaxation-modes} provides a reference for the shift and broadening induced by anharmonicity. The main response shifts towards higher reference frequencies as $\xi$ increases, consistent with this relation. At large $\xi$, the limited number of bound levels restricts repeated uptake, while the response broadens and the single-transition picture becomes less adequate.

Dissociation also becomes appreciable at strong anharmonicity. At $300$ K and $t=800$ fs, the absorbed probability $1-\lVert\Psi(t)\rVert^2$ remains below $0.05$ for the sampled $\xi<0.30$, but reaches approximately $0.68$ at $\xi=0.50$. The normalized observables at large $\xi$ therefore characterize a surviving component whose weight has decreased substantially. The zero-temperature series includes physical CAPs on the same selected modes, with the same onsets and strengths. At $\xi=0.50$ and $800$ fs its absorbed probability is approximately $0.62$, compared with the above mentioned $0.68$ at $300$ K, whereas the corresponding normalized shared energies are approximately $0.131$ and $0.133$ eV. Thus close agreement of the normalized subsystem energies does not imply equal dissociation probabilities.  %The grid differences documented in the Supplementary Material should also be kept in mind when interpreting small residual temperature differences,  
Actually, the larger value of the $T=300$ K calculation suggests a thermal contribution to the interaction-induced dissociation.

In addition to the main excitation band, the mode maps show a broad response near the lowest reference frequencies and isolated peaks below $\omega_{\rm res}$, most clearly resolved in the late-time averages reported in the bottom row of Fig.~\ref{fig:relaxation-modes}. The low-frequency background may contain a contribution from bath displacement and reorganization. The isolated peaks are already present at $0$ K and hence do not require initial thermal excitation. They are likely due to system-induced multi-quantum excitation. The nonlinear interaction in each bath coordinate allows transitions beyond adjacent vibrational levels, making two-phonon (or higher) absorption plausible. Subsystem-mediated processes involving combinations of different modes are also possible, even with no direct coupling between bath modes as in our model. The mode maps alone, however, cannot distinguish these mechanisms or establish a specific resonance assignment.

%; they are consistent with system-induced excitation involving combinations of bath modes, although the maps alone do not establish a specific combination resonance. In particular, given the absence of the intermode coupling in our model and the non-linearity of the coupling in the bath coordinates, the possibility of two- or even three-phonon absorption appears plausible.

\subsection{Damped oscillations}
\begin{figure*}[t]
\centering
\includegraphics[width=0.8\textwidth]{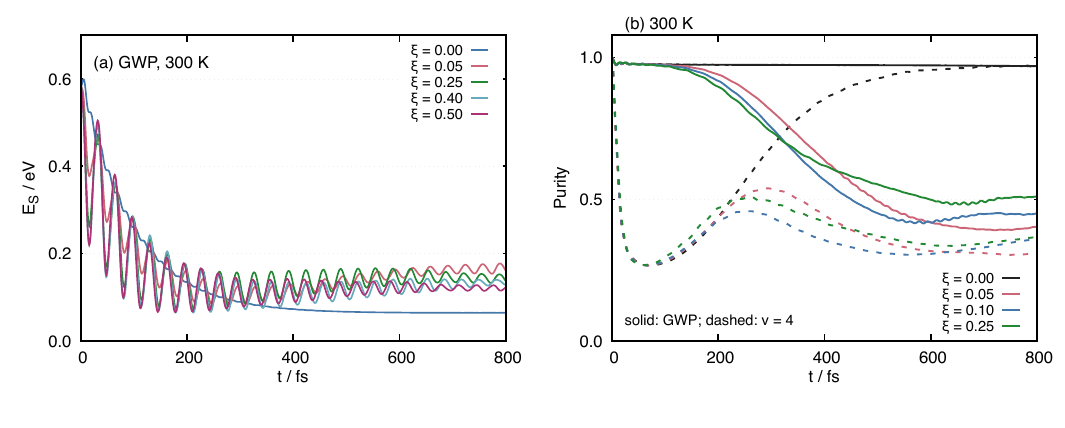}
\caption{Gaussian-wavepacket relaxation at $300$ K. (a) Bare subsystem energy for selected anharmonicities, color coded as in Fig.~\ref{fig:relaxation-bare-energy}. (b) Subsystem purity for Gaussian (solid) and $v=4$ eigenfunction (dashed) preparation at matched $\xi$, including the harmonic reference $\xi=0.00$.}
\label{fig:gwp-dynamics}
\end{figure*}

A localized Gaussian preparation probes sensitivity to the initial subsystem phase-space distribution. We write it as
\begin{equation}
 \psi_S(s)=(2\pi\sigma^2)^{-1/4}
 \exp\left[-\frac{(s-s_c)^2}{4\sigma^2}
 +\frac{i p_c(s-s_c)}{\hbar}\right].
\end{equation}
with $p_c=0$ and the width $\sigma$ matching the harmonic ground-state width. The displacement $s_c$ was set to give approximately the same average energy as in the previous problem ($0.581$ eV compared to the $0.558$ eV of the $v=4$ case). This wavepacket contains coherences in the subsystem energy basis and, in the absence of interaction with the bath, would undergo coherent, undamped oscillations without distorting. 

\begin{figure}[!t]
\centering
\includegraphics[width=\columnwidth]{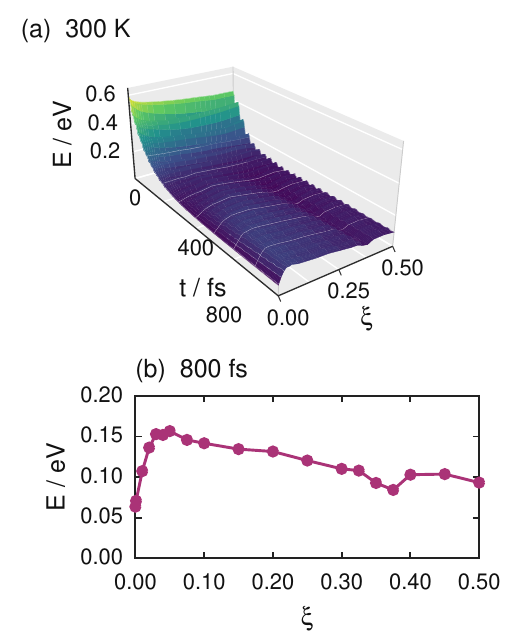}
\caption{Interaction-corrected subsystem energy $E_S^\textrm{int}$ for Gaussian preparation at $300$ K. (a) Energy decay vs. anharmonicity as in Fig.~\ref{fig:relaxation-energy}. (b) Cut at $800$ fs; symbols mark every sampled $\xi$.}
\label{fig:gwp-shared-energy}
\end{figure}

The Gaussian preparation also departs strongly from harmonic-bath relaxation at $300$ K. Anharmonic baths retain appreciable excitation and produce larger bare-energy oscillations than in the $v=4$ preparation [Fig.~\ref{fig:gwp-dynamics}(a)]. Their reduction upon including half of the interaction energy again points to a substantial reversible contribution to the energy exchange. Despite the different transient dynamics, the shared-energy cut retains a similar dependence on anharmonicity, including a minimum slightly below $\xi=0.4$ [Fig.~\ref{fig:gwp-shared-energy}(b)].

The purity remains close to unity for longer after Gaussian preparation than after eigenstate preparation, most clearly at small $\xi$ [Fig.~\ref{fig:gwp-dynamics}(b)]. For example, at $\xi=0.05$ and $100$ fs the purities are approximately $0.974$ and $0.282$, respectively. This indicates slower development of subsystem mixedness for the Gaussian initial state, although purity alone does not measure coherence in the energy basis. %The preparations have similar but unequal mean energies, so their comparison also includes differences in energy and initial level populations.

Several features agree qualitatively with the earlier Morse-bath study of L\'opez-L\'opez, Martinazzo, and Nest~\cite{MorseBath2011}: increasing anharmonicity produces stronger bare-energy oscillations, these oscillations involve the interaction energy, and they are more pronounced for a displaced Gaussian than for an eigenstate preparation. That work also connected thermal changes in relaxation to the occupation-dependent Morse transition frequencies. Quantitative agreement, or a common monotonic trend in relaxation, is not expected: the calculations differ in coupling strength, counterterm, bath preparation, anharmonicity range and treatment of dissociative modes. Indeed, the earlier study found enhanced relaxation for its Ohmic bath but suppression for a narrow Lorentzian bath. The common picture is that anharmonicity and temperature modify the bath transitions available for energy exchange; the resulting relaxation depends on the bath spectrum and preparation.

We close this section by noting that the GWP results provide a particularly useful test of the thermalization protocol because discrepancies in initial bath preparation can change subsequent energy exchange even when the initial subsystem state is identical. 

\section{Decoherence}

Localized states are robust in an environment that measures the position, as it happens with coordinate-dependent coupling. 
When preparing the system as a superposition of spatially localized states the interaction with the environment, and 
the ensuing system-bath entanglement, cause the disappearance of coherence and the emergence of the robust solutions \cite{Schlosshauer2007}. 
The phenomenon is central to the quantum-classical transition, and can be characterized by a decoherence coefficient 
$\Gamma_\textrm{coh}$ that measures the rate of `coherence decay'.
For the spatial coherence problem of interest here, two localized wavepackets $\ket{R}$ and $\ket{L}$ separated by a distance 
$\delta$ are superimposed in the initial state
\begin{equation}
    \ket{\psi_S}=c_R\ket{R}+c_L\ket{L}
\end{equation}
and allowed to evolve under coupling to the bath,
\begin{equation}
    \ket{\Psi(t)}=c_R\ket{\Psi_R(t)}+c_L\ket{\Psi_L(t)}
\end{equation}
Here, 
\begin{equation}
    \ket{\Psi_X(t)}=\exp\left(-\frac{i}{\hbar}H_\textrm{TFD}t\right)\ket{X}\ket{0(\beta)}
\end{equation}
for $X=R,L$ is the localized thermofield wavefunction at time $t$ and inverse temperature $\beta$ (noticing that the initially 
pure subsystem state saves us from doubling the system variables). The reduced density operator of the system then follows as
\begin{align}
    \rho_S(t) = &\textrm{Tr}_{B,\widetilde B}\left(\ket{\Psi(t)}\bra{\Psi(t)}\right)=\nonumber\\
    =&|c_R|^2\rho^{(R)}_S(t)+|c_L|^2\rho^{(L)}_S(t)+\nonumber\\
     & + c_R c_L^*\,\eta_{RL}(t) + \textrm{h.c}. 
\end{align}
where the trace runs over the physical and tilde bath modes, $\rho_S^{(X)}(t)$ are the localized contributions 
and the cross operator reads as
\begin{equation}
 \eta_{RL}(t)=\textrm{Tr}_{B,\widetilde B}\left(\ket{\Psi_R(t)}\bra{\Psi_L(t)}\right)
\end{equation}
Clearly, it is the presence of the latter that signals coherence in the subsystem state, 
in a way that does not depend on the specific initial superposition.  
It is then expedient to introduce the coherence norm \cite{StrunzHaake2003,StrunzHaakeBraun2003}
\begin{equation}
     n_{\rm coh}(t)=\textrm{Tr}_S[\eta_{RL}(t)\eta_{RL}^\dagger(t)]\label{eq:cohnorm}
\end{equation}
as a measure of the amount of coherence at time $t$. 

The cross operator can be reconstructed from four ordinary density matrices. To this end 
one can propagate the superposition vectors
\begin{equation}
\psi_\eta=\frac{1}{\sqrt{2}}\left(\ket{R}+\eta\ket{L}\right)    
\end{equation}
for $\eta\in\{1,-1,i,-i\}$. Denoting their (unnormalized) reduced operators by $\rho_\eta^{u}(t)$, by linearity we obtain \footnote{If the initial superpositions are normalized separately, the corresponding reduced matrices must first be multiplied by $N_\eta=1+\operatorname{Re}[\eta\langle R|L\rangle]$.}
\begin{equation}
 \eta_{RL}(t)=\frac12\sum_{\eta=1,-1,i,-i}\eta\rho_\eta^{u}(t).
 \label{eq:fourphase}
\end{equation}
Therefore, four wavepacket propagations for the basic superpositions are enough to compute the coherence norm 
and extract the decoherence rate from its decay. We opted for fitting an exponential envelope 
with an oscillatory contribution. Purity $\mathcal{P}=\textrm{Tr}\rho_S^2$ (for each $\rho_\eta$)
provides a complementary diagnostic but reflects both the mixedness of the individual wavepacket components and the coherence between them.
In fact, in the long time limit where the system approaches equilibrium,
$\mathcal{P}$ tends to its finite equilibrium value, while the coherence norm of Eq.~\eqref{eq:cohnorm} 
vanishes for well separated initial states. 
Indeed, if the reduced dynamics relaxes every initial subsystem state to the
same equilibrium state $\rho_{\mathrm{eq}}$, linearity of the dynamical map \cite{Breuer2007} implies
$\eta_{RL}(t)\to\braket{L|R}\rho_{\mathrm{eq}}$, where the prefactor follows
from trace conservation
\begin{equation}
\textrm{Tr}_S\, \eta_{RL}(t)=\braket{\Psi_L(t)|\Psi_R(t)}=\braket{L|R}\nonumber
\end{equation}
Consequently, for the coherence norm
\begin{equation}
 n_{\rm coh}(t)\rightarrow |\braket{L|R}|^2\operatorname{Tr}_S(\rho_{\mathrm{eq}}^2)\nonumber   
\end{equation}
which is negligible for well-separated wavepackets.

\subsection{Spatial superpositions in a harmonic system}
\begin{figure}[t]
 \centering
 \includegraphics[width=\columnwidth]{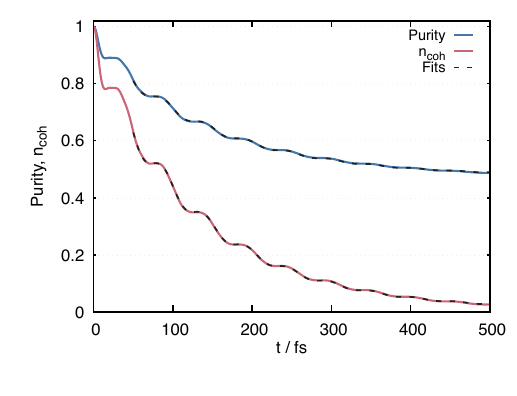}
 \caption{Coherence norm and phase-averaged purity for a spatial superposition with reduced separation $\lambda=10$, in the low-frequency subsystem ($\omega_0\simeq316.2\,\mathrm{cm}^{-1}$, $\gamma^{-1}=5$ ps), coupled to a harmonic bath ($\xi=0.00$) at $300$ K. Purity is averaged over the four phase preparations used to reconstruct the coherence operator. Colored solid curves show the dynamics; black dashed curves show damped oscillatory fits over $50$--$500$ fs, using Eq.~\eqref{eq:coherence_fit} and an analogous expression for purity.}
 \label{fig:decoherence_dynamics}
\end{figure}
\begin{figure*}[!t]
 \centering
 \includegraphics[width=0.9\textwidth]{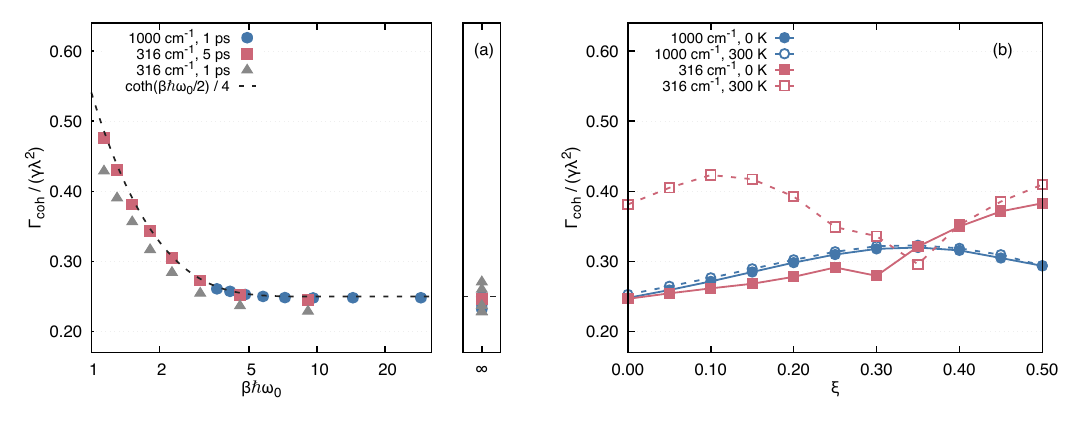}
 \caption{Scaled coherence-norm decay rates. (a) Harmonic-bath results versus $\beta\hbar\omega_0$ on a logarithmic axis, combining the temperature scans ($\lambda=10$, $T=0$--$400$ K) and some zero-temperature separation scans ($\lambda=8$--$16$). Legend entries give the subsystem frequency and relaxation time $\gamma^{-1}$. The dashed curve is $\frac14\coth(\beta\hbar\omega_0/2)$ [Eq.~\eqref{eq:scaling}]; the stronger-coupling low-frequency series shows larger deviations. The zero-temperature results ($\beta\to\infty$) are shown in a separate strip marked $\infty$. (b) Rates versus bath anharmonicity at $\lambda=10$ for the high-frequency ($\gamma^{-1}=1$ ps) and low-frequency, weak-coupling ($\gamma^{-1}=5$ ps) models. Filled symbols and solid lines denote $0$ K; open symbols and dashed lines denote $300$ K. Lines between symbols guide the eye. All rates are extracted from coherence-norm fits, by fitting over $10$--$200$ fs for the high-frequency model and $50$--$500$ fs for the low-frequency model.}
 \label{fig:decoherence_rates}
\end{figure*}

For a harmonic subsystem, we conveniently use Gaussian coherent states centered 
at $s=\pm\delta/2$ with zero mean momentum. Denoting the characteristic HO length with $\ell=\sqrt{\hbar/(2m\omega_0)}$, 
the spatial separation in reduced units defines $\lambda=\delta/\ell$. 
The overlap of the Gaussian components is $e^{-\lambda^2/8}$ 
and is about $3\times10^{-4}$ at the smallest separation considered ($\lambda=8$), 
so that we can safely assume orthogonal localized states.

We investigated two different models. The first is a high-frequency model with $\omega_0=1000\,\mathrm{cm}^{-1}$ and hydrogen mass, 
as in Sec. \ref{sec:vibrational relaxation}. In a second, lower-frequency model, we increased the mass tenfold at fixed force constant, 
giving $\omega_0\simeq316.2\,\mathrm{cm}^{-1}$. The relaxation times were set to be 1 ps in the first case, and 5 ps in the second, 
resulting in a reasonable time scale for the decoherence process. The harmonic-bath comparison also includes the low-frequency model with $\gamma^{-1}=1$ ps. 

For a harmonic environment, the expected weak-coupling scaling \cite{Burghardt2003} 
combines a quadratic separation dependence with the thermal oscillator factor $\coth(\beta\hbar\omega_0/2)$
\begin{equation}
 \frac{\Gamma_{\rm coh}}{\gamma}
 \simeq\frac{\lambda^2}{4}\coth\left(\frac{\beta\hbar\omega_0}{2}\right).
 \label{eq:scaling}
\end{equation}
where $\Gamma_{\rm coh}/\gamma$ is the decoherence ratio. The envelope rate is extracted from a damped oscillatory fit to the coherence norm of Eq.~\eqref{eq:cohnorm},
\begin{equation}
n_{\rm coh}(t)=n_{\rm coh}^{\infty}+e^{-\Gamma_{\rm coh}\tau}g(\tau),
\qquad \tau=t-t_{\rm ref},
\label{eq:coherence_fit}
\end{equation}
with $g(\tau)=A+B\cos(\Omega\tau)+C\sin(\Omega\tau)$ and $t_{\rm ref}$ the beginning of the fit interval. The oscillatory term reflects the localized wavepacket motion, with fitted $\Omega\approx2\omega_0$: the two components oscillate in antiphase and exchange positions every half-period.

Figure~\ref{fig:decoherence_dynamics} shows a representative harmonic-bath calculation together with fits of the coherence norm and the purity averaged over the four phase preparations. The coherence norm decays towards zero, whereas the purity remains finite. The fitted rates for this example are $\Gamma_{\rm coh}=7.63\times10^{-3}\,\mathrm{fs}^{-1}$ and $\Gamma_{\rm pur}=7.03\times10^{-3}\,\mathrm{fs}^{-1}$.

The harmonic-bath results in Fig.~\ref{fig:decoherence_rates}(a) combine the separation and temperature dependence in scaled units. They approximately follow Eq.~\eqref{eq:scaling}, with larger deviations for the more strongly coupled low-frequency model. At high subsystem frequency, the thermal factor changes little over the tested temperatures, whereas the low-frequency oscillator samples a more temperature-sensitive regime. The improved agreement at weaker coupling is consistent with the time-scale separation underlying the scaling relation.

% Updated 7 November 2026: completed decoherence series; see revisions/decoherence_update_2026-11-07.
Figure~\ref{fig:decoherence_rates}(b) shows contrasting responses to bath anharmonicity in the two subsystem-frequency regimes. For the high-frequency model, the $0$ and $300$ K rates differ by at most $\sim2\%$ at each sampled $\xi$. Their dependence on anharmonicity is broad: the rates rise from the harmonic limit to a maximum near $\xi=0.35$, %about $28$--$29\%$
$\sim30\%$ above the respective harmonic values, and then decrease. Thus the temperature effect is weak, while the anharmonicity effect remains appreciable. The low-frequency model is more sensitive to temperature and has a different shape. At zero temperature, its rate increases overall by more than $50\%$ %about $55\%$ 
between $\xi=0$ and $0.50$, with a shallow local dip near $\xi=0.30$. At $300$ K, the rate first rises, then develops a pronounced dip around $\xi=0.30$--$0.35$ before increasing again. The sampled minimum at $\xi=0.35$ is $5.92\times10^{-3}\,\mathrm{fs}^{-1}$, approximately $20\%$ %$22\%$ 
below the harmonic-bath value and $30\%$ below the rate at $\xi=0.10$. The dip survives changes of the fit interval, although fixing the asymptotic offset $n_{\rm coh}^{\infty}$ in Eq. (\ref{eq:coherence_fit}) to zero shifts the sampled minimum to $\xi=0.30$ (Supplementary Material). %Its precise location and depth should therefore not be assigned more significance than the finite-time fit supports.

%The low-frequency trough recalls the non-monotonic structure of the vibrational-relaxation cuts. However, a minimum of the residual energy means enhanced relaxation, whereas a minimum of $\Gamma_{\rm coh}$ means slower coherence decay. Moreover, the subsystem frequencies, preparations and coupling strengths differ between these calculations. The harmonic decoherence ratio [Eq.~\eqref{eq:scaling}] motivates a possible connection through the bath response governing dissipation: anharmonicity changes the transition frequencies, available levels and coupling matrix elements, and can thereby alter both processes. It does not establish the same scaling with a single effective damping rate for an anharmonic bath.

%A related possibility is that the increasingly sparse bound-state structure contributes to the observed changes. The Morse bound-state condition, $n+\tfrac12<1/\xi$, gives three bound levels for $2/7\leq\xi<2/5$ and only two for $2/5\leq\xi\leq0.50$. Thus the trough occurs in a few-level regime, approaching a two-level bound sector. This is suggestive of spin-bath-like behavior, but the present model also retains dissociative states and absorbing boundaries; it is not a strict spin bath. Identifying the mechanism would require transition-resolved bath diagnostics or a matched few-level comparison.

The low-frequency dip superficially resembles the non-monotonic structure of the vibrational-relaxation cuts, but this resemblance can be misleading. A minimum of the residual energy indicates enhanced relaxation, whereas a minimum of $\Gamma_{\rm coh}$ indicates slower coherence decay. The similar shapes therefore do not, by themselves, support a common dependence on an effective dissipation rate. Moreover, the subsystem frequencies, preparations and coupling strengths differ between these calculations. One concrete possibility is that the increasingly sparse bound-state structure contributes to the decoherence-rate dip. The Morse bound-state condition, $n+\tfrac12<1/\xi$, gives three bound levels for $2/7\leq\xi<2/5$ and only two for $2/5\leq\xi\leq0.50$. The dip thus occurs in a few-level regime, approaching a two-level bound sector. This raises the possibility of spin-bath-like behavior, although the present model also retains dissociative states and absorbing boundaries and is therefore not a strict spin bath. Further analysis, including transition-resolved bath diagnostics or comparisons with matched few-level models, would be needed to establish whether this reduction of the bound-state manifold causes the dip or whether other features of the anharmonic bath response dominate.

\section{Summary and Conclusions}

We have shown that separating thermal-state preparation from the Bogoliubov change of representation provides a practical route to thermofield simulations of anharmonic environments. Converting a harmonic-reference thermal state into an anharmonic target state through reference thermalization can amplify errors in high-energy auxiliary components, with increasingly severe conditioning problems as the primitive basis is enlarged. We found that the resulting preparation errors can produce rapidly growing bath occupations and unstable subsequent dynamics, even when subsystem observables initially appear well behaved. To avoid this ill-conditioned conversion, we developed two direct BT procedures (Secs.~\ref{sec:direct pair-construction}--\ref{sec:direct route to BT}). We prepare the thermal amplitude in the bare representation and subsequently apply finite-basis unitaries to the physical--auxiliary mode pairs (AA scheme); for independent modes, we can also construct the thermal amplitude directly (DC scheme). Both procedures fit naturally into an ML-MCTDH representation and yield stable finite-basis convergence in the reported preparation tests. For dissociative environments, which additionally require a specified initial ensemble, we introduced CAP-filtered preparation as a practical means of selecting a metastable ensemble dominated by bound states.

We applied the resulting protocol to model calculations involving a Morse bath with a tunable anharmonicity parameter $\xi$. The relaxation results show that bath anharmonicity changes both the transient energy exchange and the residual subsystem excitation. The late-time energy depends strongly and non-monotonically on $\xi$, with closely similar structure at $0$ and $300$ K and a local region of enhanced relaxation near $\xi=0.375$. The comparatively small temperature dependence of the residual subsystem energy is consistent with the high frequencies of the bath transitions resonant with the subsystem. Differences remain in the absorbed probability, even where observables conditioned on survival nearly coincide. The mode response shifts consistently with the anharmonic transition frequencies and develops isolated lower-frequency features suggestive of combination processes. Gaussian preparation produces more pronounced energy oscillations while preserving high purity for longer than eigenstate preparation. The oscillations and their sensitivity to preparation agree qualitatively with the earlier Morse-bath results~\cite{MorseBath2011}, although neither study supports a simple rule explaining the role of  anharmonicity in relaxation. One cannot exclude that transitions between regular and chaotic regimes in the corresponding classical phase space of the driven anharmonic bath oscillators \cite{Wiggins2019,Heller2018} influence energy transfer between the subsystem and bath.

Spatial-superposition decoherence provides a complementary measure of the bath response. Harmonic-bath rates approximately recover the expected quadratic separation dependence and thermal scaling, with larger deviations at stronger coupling. The high-frequency rates depend weakly on temperature and show a broad maximum with anharmonicity. In the low-frequency model, the rate increases overall with anharmonicity at zero temperature, whereas at $300$ K it exhibits a pronounced dip near $\xi=0.30$--$0.35$. Its resemblance to the relaxation cuts does not establish a common dependence on an effective dissipation rate, as their minima indicate enhanced energy relaxation and slower coherence decay, respectively. Resonance effects may contribute to both patterns, and their influence need not be captured by a single damping rate or by the harmonic-bath decoherence ratio. Establishing the origin of these features, including any contribution from the increasingly sparse bound-state manifold and the abovementioned regular-chaotic transition in the corresponding classical dynamics, requires further analysis.

The main outcome of the present work is methodological and it is a practical route to preparing anharmonic thermal vacua in the Bogoliubov picture while avoiding the ill-conditioned conversion from a harmonic reference. The protocol also separates thermal preparation from the treatment of dissociation, making the choice of initial ensemble and absorbing boundaries explicit. Beyond the independent Morse modes considered here, this separation offers a framework for thermofield simulations of more general anharmonic environments. Anharmonic oscillator models can provide a significantly more realistic description of chemical or biological environments than the standard harmonic oscillator bath description \cite{Makri1994AnharmonicBath,Wang2004AnharmonicBath,Makri2024AnharmonicBath,Rhee2025AnharmonicBath}. 

A further opportunity, not explored in this work, is to exploit the compact transformed wavefunction to reduce the primitive basis used for subsequent propagation. Such a reduction could follow the finite-basis Bogoliubov transformation in a sufficiently converged preparation basis, with the accuracy of the reduced representation assessed independently.

\section*{Acknowledgments}
The authors acknowledge the use of OpenAI Codex for assistance with checking numerical input settings and simulation outputs, developing and reviewing Python analysis scripts, preparing figures from simulation data, and drafting the numerical documentation in the Supplementary Material. All AI-assisted outputs were critically reviewed by the authors, who retain full responsibility for the calculations, interpretation, and final manuscript.

%\section*{Data availability}
%\checkme{Assign a public repository and versioned archive for the final inputs, analysis scripts, and figure data.}

%\bibliographystyle{apsrev4-2}
% BEGIN EMBEDDED BIBLIOGRAPHY
%apsrev4-2.bst 2019-01-14 (MD) hand-edited version of apsrev4-1.bst
%Control: key (0)
%Control: author (8) initials jnrlst
%Control: editor formatted (1) identically to author
%Control: production of article title (0) allowed
%Control: page (0) single
%Control: year (1) truncated
%Control: production of eprint (0) enabled
%

% END EMBEDDED BIBLIOGRAPHY

% BEGIN EMBEDDED SUPPLEMENTARY MATERIAL
\clearpage
\onecolumngrid
\setcounter{section}{0}
\setcounter{subsection}{0}
\setcounter{equation}{0}
\setcounter{figure}{0}
\setcounter{table}{0}
\renewcommand{\thesection}{S\arabic{section}}
\renewcommand{\thesubsection}{\thesection.\arabic{subsection}}
\renewcommand{\theequation}{S\arabic{equation}}
\renewcommand{\thefigure}{S\arabic{figure}}
\renewcommand{\thetable}{S\arabic{table}}
\renewcommand{\theHsection}{supp.\arabic{section}}
\renewcommand{\theHequation}{supp.\arabic{equation}}
\renewcommand{\theHfigure}{supp.\arabic{figure}}
\renewcommand{\theHtable}{supp.\arabic{table}}
\newcommand{\Tr}{\operatorname{Tr}}
\newcommand{\nint}{\operatorname{nint}}
\begin{center}
{\Large\bfseries Supplementary Material for\\[4pt]
Bogoliubov-picture thermofield dynamics\\in anharmonic environments}\par
\vspace{8pt}
Rocco Martinazzo, Bartosz B{\l}asiak and Irene Burghardt
\end{center}

\section{ML-MCTDH workflow}
\begin{figure}[p]
\centering
\includegraphics[width=0.96\linewidth,height=0.85\textheight,keepaspectratio]{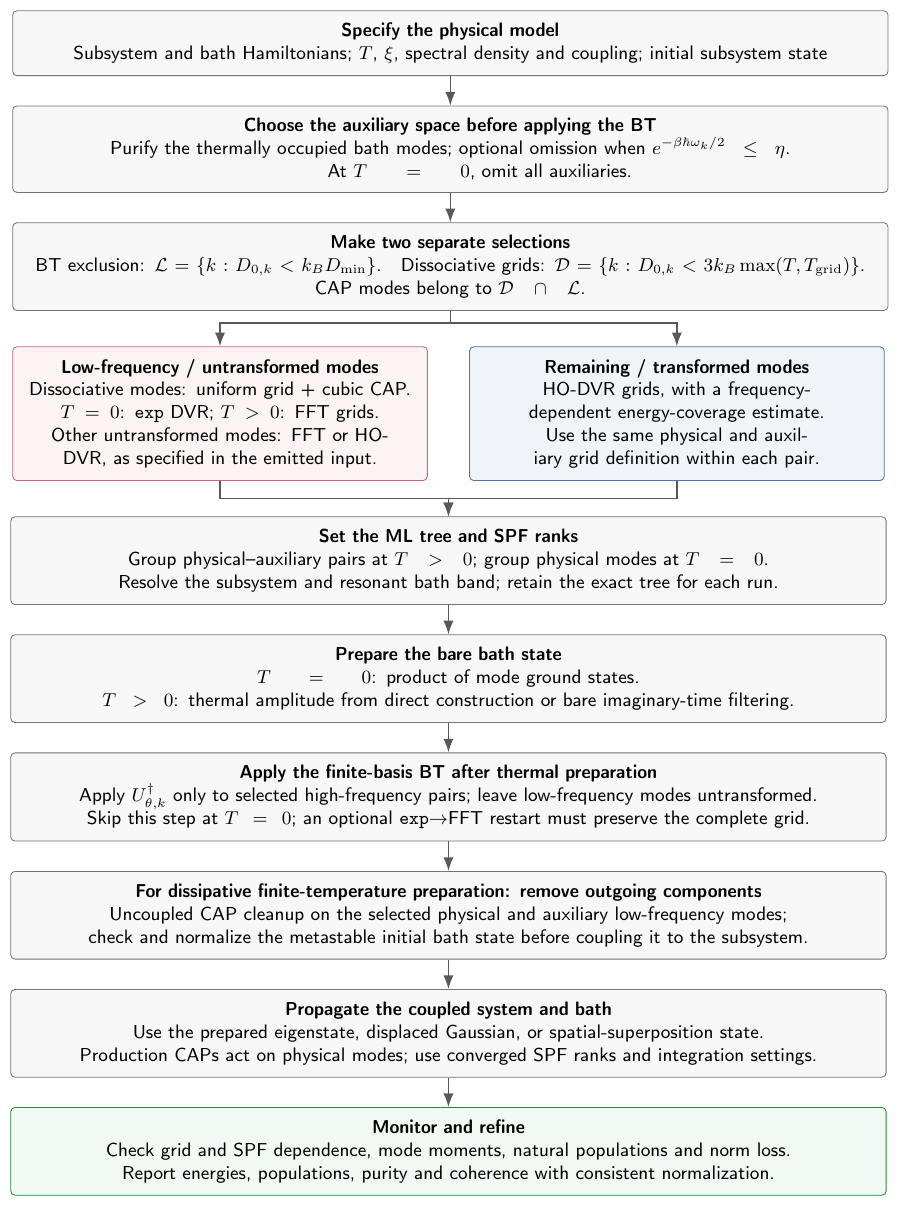}
\caption{Workflow for thermal preparation and ML-MCTDH dynamics. The two mode-selection criteria have different purposes: exclusion from the BT and provision of dissociative grids. The actual settings for the calculations described in the main text are listed separately in Table~\ref{supp:tab:families}. Auxiliary omission is optional, and all auxiliaries are absent at zero temperature. 
%Grid, SPF and CAP convergence remain independent checks.
}
\label{supp:fig:workflow}
\end{figure}

Figure~\ref{supp:fig:workflow} summarizes the workflow used to prepare and propagate the thermofield wavefunction. We developed Fortran and Python tools to automate the generation of MCTDH inputs and the application of finite-basis Bogoliubov transformations. The Fortran input generators construct the bath discretization, Hamiltonian operators, primitive bases and initial-state specifications from the physical model parameters and numerical coverage criteria. Python tools construct the physical--auxiliary pair unitaries and generate the staged inputs needed to apply them. The workflow distinguishes the selection of modes for the Bogoliubov transformation from the provision of extended grids and complex absorbing potentials for potentially dissociative modes. The generated inputs provide starting settings; primitive-grid, absorber and multilayer-basis convergence require separate checks.

At finite temperature, we first prepare the thermal amplitude in the bare representation, either by imaginary-time propagation or, for independent bath modes, by direct construction. The Bogoliubov-transformation tools then construct finite-basis pair unitaries and generate staged inputs that apply them to the selected physical--auxiliary pairs. Where required, uncoupled CAP filtering selects the metastable initial ensemble, which is normalized before the coupled dynamics. At zero temperature, we initialize the bath in its uncoupled ground state without auxiliary coordinates. The resulting wavefunctions serve as initial states for ML-MCTDH propagation with the corresponding Hamiltonian representation. 

The following sections specify the mode-selection rules, primitive grids, absorbing potentials, multilayer trees and integration settings. 

\clearpage
\section{Scope and calculation sets}
The bath-coordinate mass is $M=1$ amu. All models contain 50 physical bath oscillators, with reference frequencies spaced by $40\,\mathrm{cm}^{-1}$ up to $2000\,\mathrm{cm}^{-1}$. The subsystem has frequency $1000\,\mathrm{cm}^{-1}$ and hydrogen mass for vibrational relaxation and for the high-frequency decoherence model. Increasing the subsystem mass tenfold at fixed force constant gives the low-frequency model, $\omega_0\simeq316.2\,\mathrm{cm}^{-1}$. The corresponding relaxation times are $\gamma^{-1}=100$ fs for vibrational relaxation, 1 ps for high-frequency decoherence, and 5 ps for the weak-coupling low-frequency model. A low-frequency harmonic reference with $\gamma^{-1}=1$ ps is also retained.

Table~\ref{supp:tab:families} reports ranges from the production inputs. %; inventories retain exact primitive definitions, CAP parameters and ML ranks. 
The 59 relaxation cases comprise 22 zero-temperature eigenstate, 18 finite-temperature eigenstate and 19 Gaussian preparations. The 75 decoherence parameter sets use four phase propagations each, giving 300 inputs. Of these, 44 parameter sets (176 phase runs) comprise the completed $\lambda=10$ anharmonicity scans for the two manuscript models at 0 and 300 K. %The remaining harmonic temperature and separation scans are retained from the existing archive; the high-frequency zero-temperature separation inputs are stored in the legacy \texttt{T-0K\_old} directory.

\begin{table}[htbp]\centering\small
\caption{Settings read from the expanded inputs for the data currently plotted. Ranges span all members of each family. The primitive dimensions refer to bath coordinates. SPF ranges refer to incoming ML branches, excluding the root and primitive dimensions. Four phase propagations are counted as one decoherence parameter set.}\label{supp:tab:families}
\setlength{\tabcolsep}{3pt}\renewcommand{\arraystretch}{1.15}
\begin{tabular}{@{}p{0.43\linewidth}rrrrrr@{}}\toprule
Family & Cases & $N_{\rm aux}$ & HO-DVR & FFT/exp & CAP modes & SPFs\\\midrule
Eigenstate, 0 K & 22 & 0 & 10--52 & 25--225 & 0--27 & 5--17 \\
Eigenstate, 300 K (Morse) & 17 & 50 & 10--27 & 16--324 & 0--27 & 5--18 \\
Eigenstate, 300 K (harmonic) & 1 & 50 & 10--52 & -- & 0 & 4--14 \\
Gaussian, 300 K (Morse) & 18 & 50 & 6--27 & 16--400 & 0--27 & 5--18 \\
Gaussian, 300 K (harmonic) & 1 & 50 & 8--52 & -- & 0 & 5--18 \\
Decoherence, high frequency & 33 & 0--41 & 8--30 & 16--225 & 0--27 & 4--12 \\
Decoherence, low frequency, weak coupling & 29 & 0--41 & 8--30 & 16--225 & 0--27 & 4--12 \\
Decoherence, low frequency, stronger coupling & 13 & 0--41 & 8--30 & -- & 0 & 4--12 \\
\bottomrule\end{tabular}
\end{table}

All finite-temperature relaxation calculations retain 50 auxiliaries, including the updated harmonic eigenstate reference. Some decoherence calculations omit thermally inactive auxiliaries. % and retain extra coordinates \texttt{dum1} and \texttt{dum2}, which are excluded from bath counts.

The subsystem uses 64 FFT points: $[-3,3]a_0$ for relaxation, $[-4,4]a_0$ for high-frequency decoherence and $[-2.5,2.5]a_0$ for low-frequency decoherence, with linear endpoint convention; uniform bath grids are periodic. The completed zero-temperature series uses uniform grids and CAPs for dissociative modes, and HO-DVR otherwise. At common sampled anharmonicities the physical CAP modes, onsets and strengths match those at 300 K. The zero-temperature runs use a slightly wider left-boundary rule than the 300 K runs, and their odd uniform-grid dimensions reflect exponential-DVR initialization. The right endpoint can also differ where a minimum grid dimension extends the CAP interval. %At $\xi=0.010$ and $0.025$, $q_1$ uses HO-DVR at zero temperature and FFT at 300 K; at $\xi=0.050$, $q_1$ has a CAP at both temperatures, while the nonabsorbing $q_2$ uses HO-DVR at zero temperature and FFT at 300 K. All these differences are recorded in the inventory; identical grids across temperatures are not assumed.

\section{Mode selection and primitive bases}
\subsection{Thermal representation and dissociative grids}
The auxiliary space is chosen before applying the thermal Bogoliubov transformation (BT). An optional selection retains modes satisfying
\begin{equation}
 e^{-\beta\hbar\omega_k/2}>\eta.
\end{equation}
At zero temperature, the bath is initialized in a pure product state and requires no auxiliaries. At finite temperature, low-frequency Morse modes can require a different primitive representation from the higher-frequency modes. Two thresholds serve different purposes and must be kept separate.

With $\alpha_k=\sqrt{M\omega_k\xi/\hbar}$, the Morse depth and the dissociation gap above the ground state are
\begin{equation}
 D_{e,k}=\frac{\hbar\omega_k}{2\xi},\qquad
 D_{0,k}=D_{e,k}(1-\xi/2)^2.
\end{equation}
The input parameter $D_{\min}$, in kelvin, defines the low-frequency set excluded from the BT,
\begin{equation}
 \mathcal L=\{k:D_{0,k}<k_B D_{\min}\}.
\end{equation}
For the retained Morse model these modes remain Morse oscillators; a replacement with a torsional potential is also possible but not used in the present study. A value in the range $D_{\min}=800-900$ K was found to be appropriate for both relaxation and decoherence and used throughout. %The emitted input, including any later edits, determines the actual numerical setting.

The input generator introduces a separate empirical grid-selection scale $T_{\rm grid}$:
\begin{equation}
 T_* = \max(T,T_{\rm grid}),\qquad
 \mathcal D=\{k:D_{0,k}<3k_BT_*\}.
 \label{supp:eq:gridcriterion}
\end{equation}
At finite temperature, the implementation restricts CAPs to $\mathcal D\cap\mathcal L$; a mode outside $\mathcal L$ is not silently removed from the BT. At zero temperature, $\mathcal L$ is reset to $\mathcal D$. The input therefore specifies $D_{\min}$, $T_{\rm grid}$ and the representation flags independently. Equation~\eqref{supp:eq:gridcriterion} is a numerical coverage rule, not a prediction of the dissociation probability in the coupled dynamics. Its purpose is to retain dissociative grids even when the initial thermal population is negligible.

\subsection{HO-DVR coverage}
The standard current HO-DVR rule uses a frequency-dependent coverage estimate, with a mode-dependent frequency cutoff, 
\begin{equation}
 \omega^c_k=\begin{cases}
 \Omega,&\omega_k<\omega_5,\\
 \Omega\left[1+3\sqrt{(\omega_k-\omega_5)/(\Omega-\omega_5)}\right],&\omega_k\geq\omega_5,
 \end{cases}
\end{equation}
where $\Omega$ is the frequency cutoff of the bath. The corresponding number of primitive functions is
\begin{equation}
 N_k^{\rm HO}=\max\left[10,\;2+\max\left\{\nint(\omega^c_k/\omega_k),
 \nint(\omega^c_{N}/\Omega)\right\}\right].
 \label{supp:eq:ho}
\end{equation}
This provides additional excitation capacity and two extra functions beyond the estimated energy coverage. It is a starting rule, not a convergence guarantee. The branch for low-frequency, untransformed but nonabsorbing modes can use a different estimate based on the first nondissociative mode. %Earlier inputs also contain smaller minimum dimensions or manual reductions. These differences are recorded in Table~\ref{supp:tab:families} and in the exact grid inventory; Eq.~\eqref{supp:eq:ho} is not assigned retrospectively to every archived run.

\subsection{Uniform grids and absorbing potentials}
For a selected dissociative mode, the input generator targets a kinetic-energy ceiling $E_{\rm kin}=0.025$ hartree, approximately $0.680$ eV. In atomic units the target spacing and nominal endpoints are
\begin{align}
 \Delta q_{\rm target}&=\frac{\pi}{\sqrt{2ME_{\rm kin}}},\\
 q_-&=-\alpha_k^{-1}\ln\left[1+\sqrt{1+C_{\rm left}k_BT_*/D_{e,k}}\right],\\
 q_+^{(0)}&=(1+3\pi/2)\alpha_k^{-1}.
 \label{supp:eq:endpoints}
\end{align}
The left endpoint corresponds to an energy $D_{e,k}+C_{\rm left}k_BT_*$ above the well minimum, with $T_*=\max(T,T_{\rm grid})$. The completed zero-temperature cases use $C_{\rm left}=50$ and $T_{\rm grid}=300$ K; the finite-temperature relaxation inputs use $C_{\rm left}=10$. The wider zero-temperature boundary was selected after short-time grid tests with the corrected physical Hamiltonian. The scale $T_{\rm grid}$ controls numerical coverage and does not change the physical preparation temperature. The kinetic-energy ceiling controls the target spacing independently of this boundary criterion.

For FFT grids, the estimated dimension is rounded to the nearest integer with prime factors 2, 3 and 5, with at least 16 points. This rounding can be downward. If the minimum dimension requires extension of the positive endpoint, the endpoint is extended before writing the grid. The actual periodic spacing is $(q_+-q_-)/N$; thus the final dimension and endpoints, rather than the nominal kinetic ceiling alone, establish the represented grid.

At zero temperature, dissociative modes use exponential DVR (\texttt{exp}) for initialization with \texttt{eigenf}, which requires an explicit kinetic-energy matrix. Their dimensions are rounded upward to odd FFT-compatible values $N=3^a5^b$, without making the target spacing coarser. The \texttt{exp} and FFT representations can be interchanged when the complete grid is identical. FFT can be more efficient for larger grids; an optional preparation with \texttt{exp} can therefore be followed by reading its restart in a production input using FFT. This requires unchanged dimensions, endpoints and endpoint convention. A basis-type warning is expected on this specific restart.

The right-facing cubic CAP is added to the Hamiltonian as
\begin{equation}
 -iW_k(q)=-iA_k(q-q_{c,k})^3\Theta(q-q_{c,k}),\qquad
 q_{c,k}=\frac{3\pi}{2\alpha_k},\qquad A_k=0.004\,\alpha_k^{5/2},
 \label{supp:eq:cap}
\end{equation}
with the coefficient expressed in the generator's atomic-unit convention. Its nominal length is $\alpha_k^{-1}$. An endpoint extension for the minimum grid dimension increases the actual length $q_+-q_{c,k}$; it does not move the stored CAP onset. For the completed zero-temperature inputs, the wider left endpoint and the odd point count are retained in FFT production. The minimum dimension is 25, compared with 16 in the finite-temperature FFT inputs; therefore some right endpoints and actual CAP lengths also differ, even though CAP onsets and strengths agree. The strength scaling is empirical. Grid and CAP reflection checks remain separate from SPF convergence.

For example, the $T=300$ K, $\xi=0.40$ eigenstate run has 19 physical CAP modes. Mode $q_1$ uses 64 FFT points with periodic endpoints $[-5.53986,15.67003]a_0$; its CAP onset is $12.926866a_0$, its cubic coefficient is $0.000321$ in atomic units, and its actual length is $2.743164a_0$. Mode $q_{19}$ uses 16 points with endpoints $[-0.64513,4.91122]a_0$ and onset $2.965626a_0$, giving the longer actual CAP interval $1.945594a_0$. Modes $q_{20}$--$q_{50}$ use HO-DVR grids in this run.

\section{Preparation, tree structure and integration}
Figure~\ref{supp:fig:workflow} separates bare thermal preparation from the subsequent finite-basis BT on selected physical--auxiliary pairs. Independent-mode thermal amplitudes can be constructed directly or by bare imaginary-time filtering. The 300 K example uses $\beta\hbar/2=12.7305$ fs and VMF with \texttt{ABM/all = 6, 1e-10, 1e-4}; pair transformations are staged in groups of at most fifteen operators. Uncoupled CAP cleanup can act on both members of low-frequency pairs before normalization; production CAPs act on physical modes only.

\begin{turnpage}
\begin{figure}[p]
\centering
\includegraphics[width=\linewidth]{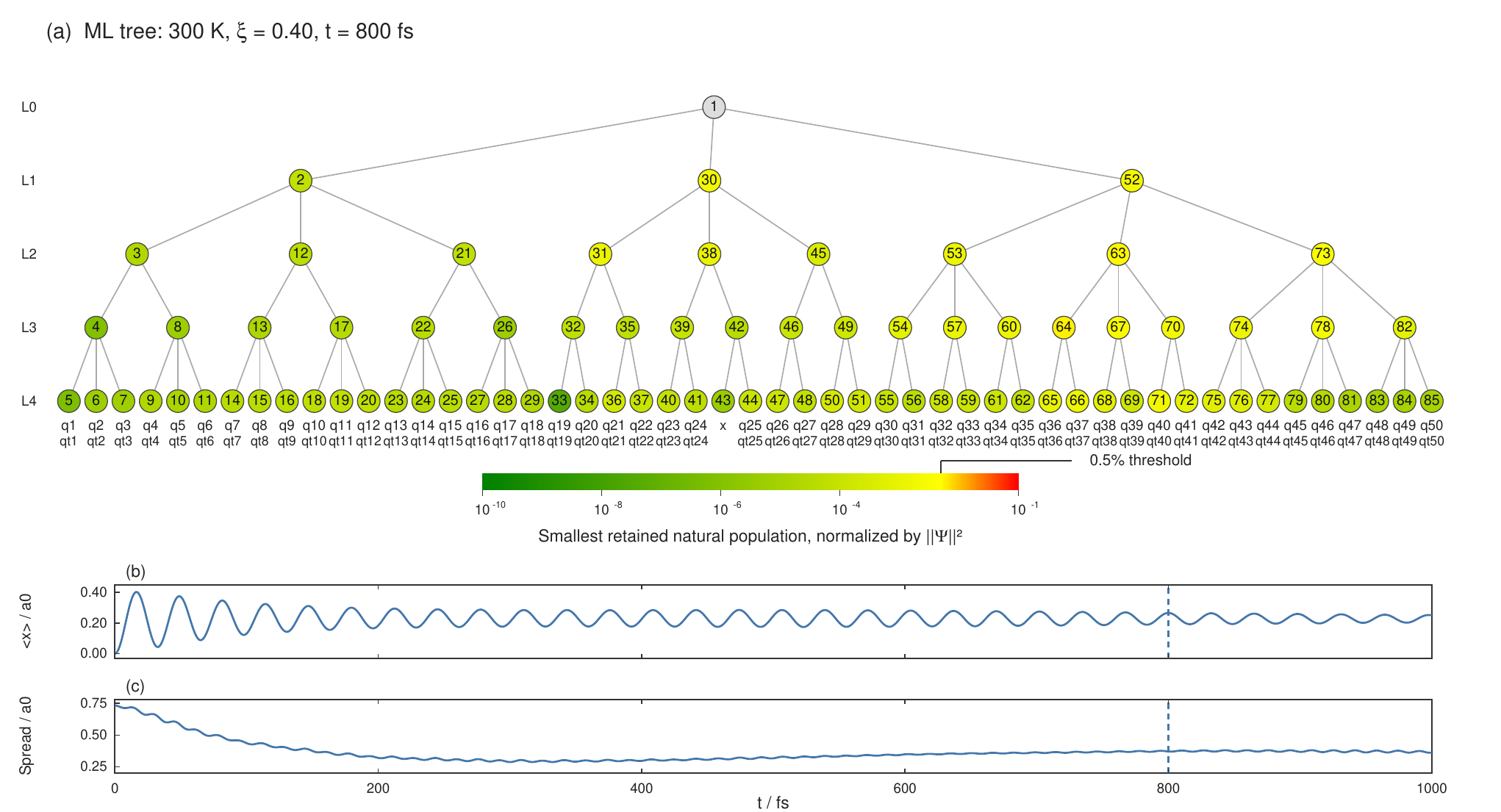}
\caption{Representative ML tree and diagnostics for eigenstate relaxation at $T=300$ K and $\xi=0.40$. (a) Incoming branches are colored by their smallest retained natural population at 800 fs, normalized by $\lVert\Psi\rVert^2$. The root is gray because it has no incoming branch. Numbers denote ML nodes, not SPF ranks; the leaf labels give grouped coordinates. The logarithmic color scale runs from $10^{-10}$ to $10^{-1}$; yellow marks the $0.5\%$ diagnostic threshold and larger values progress towards red. Values outside the scale use the endpoint colors. At 800 fs, two of 84 branches exceed $0.5\%$, with a maximum of $0.588\%$. (b,c) Subsystem mean coordinate and spread, in $a_0$; the dashed cursor marks 800 fs. This snapshot illustrates the representation and its diagnostics without asserting a universal truncation threshold or replacing observable-based convergence tests.}
\label{supp:fig:tree}
\end{figure}
\end{turnpage}

Physical--auxiliary pairs share combined modes at finite temperature; zero-temperature combined modes group physical coordinates. Figure~\ref{supp:fig:tree} documents one finite-temperature topology: 101 coordinates, 85 ML nodes, 12 SPFs on each root branch, and internal/combined-mode ranks of 5--12. Primitive leaves give coordinate labels; node numbers identify the complete rank inventory. Trees and ranks vary between runs.

All eigenstate relaxation and Gaussian runs use split CMF with \texttt{CMF/var = 0.03, 1e-8}, \texttt{RK8/def = 1e-8, 0.0, imp-ortho}, and top-node \texttt{SIL/@1 = 20, 1e-8}. The value 0.03 is the CMF update parameter; actual steps are variable. The decoherence anharmonicity scans use the same split-CMF integration settings. %The retained earlier harmonic scans include unified ML-CMF with \texttt{RK8/all = 1e-8}; the inventory distinguishes these inputs. Several small-$\xi$ Gaussian runs use outer CMF accuracy $10^{-6}$; the $\xi=0.50$ Gaussian run uses VMF with \texttt{ABM/all = 6, 1e-6, 1e-4}.

The zero-temperature relaxation series comprises 22 runs, all extending to 1000 fs and including reduced-density-matrix data. %The optimized ML trees were retained, with 47--49 nodes, incoming ranks of 5--17 and a maximum ML layer index of 4--5 across the family. %These run-specific values are listed in the inventory. 
Relaxation plots normally show 0--800 fs. Decoherence runs extend to 200 fs, or 500 fs for the weak-coupling low-frequency model. Observables are normally recorded every 1 fs.
%; fit windows are retained in the fit catalog. Integration, SPF, primitive-grid and CAP convergence are independent controls.

The natural populations printed by ML-MCTDH are eigenvalues of reduced density matrices of the full wavefunction; orthonormal SPFs do not imply that these matrices have unit trace when a CAP is present. We checked all 84 branch spectra of a representative run at 800 fs: their sums lie between 0.671406 and 0.671409, consistent with $S=\lVert\Psi\rVert^2=0.6714075$ within printing precision. This holds both at the root and at lower layers. Figure~\ref{supp:fig:tree} therefore uses relative populations $\bar p_j=p_j/S$, describing the surviving component. %Native weights remain valid absolute diagnostics; this rescaling specifies the convention used for the plotted threshold, rather than correcting the propagation.

For minimum-population diagnostics we prefer the native \texttt{lownatpop} file, whose header identifies each layer, parent node and branch. It avoids rounding and mixed formatting in \texttt{output}: decimal natural weights there are multiplied by 1000, while scientific-notation entries are unscaled, even within one spectrum. Full spectra from \texttt{output} remain necessary for purity and participation measures. The analysis supports both sources and both raw and survival-normalized conventions; for normalized spectra, $\sum_jp_j^2$ is divided by $S^2$. Small retained populations are convergence indicators, not observable error bounds; rank, tree, grid and integration checks remain independent.

For the mode maps, HO-DVR occupations are read directly from the output. FFT/\texttt{exp} modes instead print position and momentum means and standard deviations; in atomic units the same harmonic-reference occupation is reconstructed as
\begin{equation}
 \langle n_k\rangle=\tfrac12\left[M\omega_k\big((\langle q_k\rangle-q_{0,k})^2+\sigma_{q,k}^2\big)+\frac{\langle p_k\rangle^2+\sigma_{p,k}^2}{M\omega_k}\right]-\tfrac12.
\end{equation}
These printed moments are already survival-normalized and need no further rescaling. Reconstruction retains their four-decimal output precision; subtracting the initial value gives the plotted $\Delta\langle n_k\rangle$.

\subsection{Comparison of the relaxation series}
Both temperatures show the sampled local minimum of the survival-normalized shared subsystem energy near $\xi=0.375$ at 600 and 800 fs. Over the 18 common anharmonicities, the root-mean-square differences between the temperature cuts are 4.37 and 4.00 meV, respectively. At $\xi=0.50$ and 800 fs the shared energies are 0.13113 eV at zero temperature and 0.13288 eV at 300 K, whereas the surviving squared norms are 0.37815 and 0.32274. The close energy agreement therefore coexists with different absorbed fractions. These comparisons %use the final \texttt{evolve-ML1} archives only; earlier \texttt{evolve-ML1a/b} runs are excluded. They 
do not establish that the remaining cross-temperature grid differences are negligible for every bath observable. SPF convergence, primitive-grid sensitivity and CAP sensitivity are separate checks.

\subsection{Decoherence series and rate extraction}
The completed scans contain eleven anharmonicities, $\xi=0.00,0.05,\ldots,0.50$, at each temperature and subsystem frequency. At 300 K these inputs retain 31 auxiliary bath coordinates; at zero temperature none are present. The zero-temperature low-frequency coordinates use \texttt{exp} DVR in production, whereas the corresponding 300 K coordinates use FFT. The uniform-grid minimum is 25 points at zero temperature and 16 at 300 K. The nominal boundary rule uses $T_{\rm grid}=300$ K and the wider left-boundary factor 50. %Nevertheless, the primitive representations and CAP sets are not identical across temperatures: at $\xi=0.50$, for example, there are 27 physical CAP modes at zero temperature and 24 at 300 K, because the finite-temperature CAP set is restricted to the low-frequency set excluded from the BT. For $q_1$ the periodic endpoints are $[-6.97662,14.01570]a_0$ at both temperatures, and the dimensions are 75 and 64, respectively. %The inventory retains these distinctions rather than assigning a common basis to the two temperatures.

For each anharmonic 300 K phase preparation, the restart chain is bare thermal preparation, staged finite-basis BT, uncoupled CAP cleanup for 5 ps, normalization, and coupled dynamics. We checked that all 80 anharmonic phase runs read the final BT stage into cleanup and then follow this chain, with identical primitive definitions at the BT stage and in production. All 176 production phases in the four completed scans terminate normally, and all 44 archived rate files match their calculation-directory counterparts. Across all 176 phases, the largest survival-normalized minimum retained population read from \texttt{lownatpop} over the trajectories is $1.88\times10^{-4}$, below the $0.5\%$ diagnostic threshold. This is an SPF diagnostic, not a bound on errors from the primitive grids or absorbers.

The coherence norm is reconstructed from the four phase preparations, and the representative purity is their arithmetic mean. The archived damped-oscillatory fits use 10--200 fs for the high-frequency model and 50--500 fs for the low-frequency model, with the same interval at both temperatures. Table~\ref{supp:tab:decoherence-rates} lists the rates used in the manuscript figure. The harmonic temperature and separation comparisons retain their individual archived fitting windows. %Legacy harmonic aliases are excluded when a completed explicit-$\xi=0$ rerun exists, so that each physical parameter set is counted once.

As a sensitivity check, we refitted the low-frequency curves over 10--500, 50--500, 100--500, 150--500 and 50--400 fs, both with a free asymptotic offset and with that offset fixed to zero. The 300 K trough persists in all these checks: the sampled minimum occurs at $\xi=0.35$ for the free-offset fits and at $\xi=0.30$ for the zero-offset fits. At $\xi=0.35$, the archived free-offset fit gives $n_{\rm coh}^{\infty}=-0.02266$ and $\Gamma_{\rm coh}=5.922\times10^{-3}\,\mathrm{fs}^{-1}$; constraining the offset to zero gives $6.548\times10^{-3}\,\mathrm{fs}^{-1}$ over the same 50--500 fs interval. A negative fitted offset is an empirical finite-window parameter, not a physical asymptotic coherence norm. These comparisons support the non-monotonic trend without determining a unique asymptotic rate or an exact minimum. The high-frequency fits were also checked over 10--150 and 20--200 fs, with free and zero offsets. %; all sensitivity results accompany the rate inventory.

%\clearpage
\begin{table}[htbp]\centering\small
\caption{Completed decoherence rates $\Gamma_{\rm coh}$ in $10^{-3}\,\mathrm{fs}^{-1}$ for $\lambda=10$. Values are the archived free-offset fits used in the manuscript; the number of reported digits does not imply statistical uncertainty.}\label{supp:tab:decoherence-rates}
\begin{tabular}{@{}rrrrr@{}}\toprule
 & \multicolumn{2}{c}{High frequency} & \multicolumn{2}{c}{Low frequency}\\
$\xi$ & 0 K & 300 K & 0 K & 300 K\\\midrule
0.00 & 24.815 & 25.291 & 4.937 & 7.627 \\
0.05 & 25.935 & 26.464 & 5.093 & 8.102 \\
0.10 & 27.167 & 27.670 & 5.235 & 8.466 \\
0.15 & 28.508 & 28.954 & 5.366 & 8.348 \\
0.20 & 29.843 & 30.238 & 5.563 & 7.854 \\
0.25 & 30.998 & 31.385 & 5.830 & 6.982 \\
0.30 & 31.801 & 32.181 & 5.593 & 6.723 \\
0.35 & 32.001 & 32.310 & 6.424 & 5.922 \\
0.40 & 31.588 & 31.878 & 6.997 & 7.045 \\
0.45 & 30.503 & 30.964 & 7.429 & 7.708 \\
0.50 & 29.382 & 29.400 & 7.661 & 8.194 \\
\bottomrule\end{tabular}
\end{table}

The possible few-level interpretation follows from the Morse bound-state condition $n+1/2<1/\xi$: there are three bound levels for $2/7\leq\xi<2/5$ and two for $2/5\leq\xi\leq0.50$. The finite grids additionally represent dissociative states and CAP loss, so the two-bound-level regime does not define a strict spin bath. No transition-resolved attribution of the coherence-rate trough is made here.

%\section{Accompanying inventories}
%The machine-readable inventories include post-generation edits. \nolinkurl{runs.csv} records counts, dimensions, integrators and durations; \nolinkurl{primitive_grids.csv} gives coordinate definitions and CAP parameters; \nolinkurl{ml_tree_ranks.csv} gives topology and ranks; \nolinkurl{matched_grid_comparison.csv} compares physical grids and CAPs at common $\xi$; \nolinkurl{source_manifest.csv} records input checksums; and \nolinkurl{tree_diagnostic_800fs.csv} supplies the representative branch minima, with an accompanying provenance record. The completed decoherence scans additionally provide \nolinkurl{decoherence_rates_and_fits.csv}, \nolinkurl{decoherence_phase_validation.csv}, \nolinkurl{decoherence_fit_sensitivity.csv} and \nolinkurl{decoherence_source_manifest.csv}.

\clearpage

% END EMBEDDED SUPPLEMENTARY MATERIAL
\end{document}